\documentclass[11pt,a4paper]{article}
\usepackage[english]{babel}
\usepackage{amsmath,amsthm,amssymb,epsfig,latexsym}
\usepackage[numbers,square,sort&compress]{natbib}
\newcommand{\la}{u}
\newcommand{\muu}{v}
\newcommand{\lac}{u^{\scriptscriptstyle C}}
\newcommand{\lab}{u^{\scriptscriptstyle B}}
\newcommand{\muc}{v^{\scriptscriptstyle C}}
\newcommand{\mub}{v^{\scriptscriptstyle B}}
\newcommand{\as}{\lambda}
\newcommand{\bla}{\bar u}
\newcommand{\bmu}{\bar v}
\newcommand{\blac}{\bar{u}^{\scriptscriptstyle C}}
\newcommand{\blab}{\bar{u}^{\scriptscriptstyle B}}
\newcommand{\bmuc}{\bar{v}^{\scriptscriptstyle C}}
\newcommand{\bmub}{\bar{v}^{\scriptscriptstyle B}}
\newcommand{\so}{\scriptscriptstyle \rm I}
\newcommand{\st}{\scriptscriptstyle \rm I\hspace{-1pt}I}
\newcommand{\qo}{\scriptscriptstyle \rm{ I}}
\newcommand{\qt}{\scriptscriptstyle \rm I\hspace{-1pt}I}
\newcommand{\be}[1]{\begin{equation}\label{#1}}
\newcommand{\ba}[1]{\begin{multline}\label{#1}}
\newcommand{\ee}{\end{equation}}
\newcommand{\ea}{\end{eqnarray}}

\newcommand{\num}{\\\rule{0pt}{20pt}}

\newcommand{\dis}{\displaystyle}

\newcommand{\diag}{\mathop{\rm diag}}
\newcommand{\tr}{\mathop{\rm tr}}

\def\eee{{\rm e}}
\newcommand{\fj}{{\mathfrak{j}}}

\newtheorem{lemma}{Lemma}[section]

 \makeatletter
 \@addtoreset{equation}{section}
 \makeatother
 
\begin{document}

\begin{flushright}
LAPTH-029/12
\end{flushright}

\vspace{20pt}

\begin{center}
\begin{LARGE}
{\bf The algebraic Bethe ansatz for scalar products in $SU(3)$-invariant integrable models}
\end{LARGE}

\vspace{50pt}

\begin{large}
{S.~Belliard${}^a$, S.~Z.~Pakuliak${}^b$, E.~Ragoucy${}^c$, N.~A.~Slavnov${}^d$\footnote[1]{samuel.belliard@univ-montp2.fr, pakuliak@theor.jinr.ru, eric.ragoucy@lapp.in2p3.fr, nslavnov@mi.ras.ru}}
\end{large}

 \vspace{15mm}

${}^a$ {\it  Universit\'e Montpellier 2, Laboratoire Charles Coulomb,\\ UMR 5221,
F-34095 Montpellier, France}

\vspace{5mm}

${}^b$ {\it Laboratory of Theoretical Physics, JINR, 141980 Dubna, Moscow reg., Russia,\\
Moscow Institute of Physics and Technology, 141700, Dolgoprudny, Moscow reg., Russia,\\
Institute of Theoretical and Experimental Physics, 117259 Moscow, Russia}

\vspace{5mm}

${}^c$ {\it Laboratoire de Physique Th\'eorique LAPTH, CNRS and Universit\'e de Savoie,\\
BP 110, 74941 Annecy-le-Vieux Cedex, France}

\vspace{5mm}

${}^d$ {\it Steklov Mathematical Institute,
Moscow, Russia}

\end{center}

\vspace{1cm}

\begin{abstract}
We study $SU(3)$-invariant integrable models solvable by nested algebraic Bethe ansatz.
We obtain a determinant representation for the particular case of scalar products of
Bethe vectors. This representation can be used for the calculation of form factors
and correlation functions of XXX $SU(3)$-invariant Heisenberg chain.
\end{abstract}

\newpage

\tableofcontents

\section{Introduction}

In the present paper we consider the problem of calculating scalar products in the framework
of the algebraic Bethe Ansatz.

The algebraic Bethe Ansatz is a powerful method to study  quantum integrable models
\cite{FadST79,BogIK93L,FadLH96}.  It gives an effective tool for the evaluation of the spectrum of quantum Hamiltonians
\cite{KulS79,FadT79}.  The computation of  form factors of local operators
and correlation functions for many integrable models also can be performed in the framework
of the algebraic Bethe Ansatz \cite{Kor84,IzeK84}.  This last problem in various cases can be reduced to the calculation
of scalar products between off-shell Bethe vectors, that is Bethe vectors where the Bethe parameters are not required to satisfy  Bethe Ansatz
equations\footnote[1]{See section \ref{S-N} for more details on Bethe vectors.\label{ft-bethe}}.

The problem of scalar products was first considered for  $\frak{gl}_2$-based integrable models \cite{Kor82,IzeK84}.
In these works a recursive formula
for the scalar product of generic off-shell Bethe vectors was obtained (Izergin--Korepin formula).
Important progress in this study was achieved in \cite{Ize87}, where a determinant representation for the highest coefficient
of a scalar product was derived. This made the Izergin--Korepin formula explicit, however it still remained rather cumbersome for applications.
The next step was done in \cite{Sla89}, where a determinant representation for the
scalar products involving  on-shell Bethe vectors\footnote[2]{%
That is eigenvectors of the transfer-matrix. See section \ref{S-N} for more details.}
was obtained. Determinant representations for particular cases of scalar products became extremely important
after the quantum inverse scattering problem was solved \cite{KitMT99,MaiT00}. Using the explicit
solution of the inverse scattering problem one can reduce the calculation of  correlation functions
and form factors of local operators to scalar products, in which one of the vectors is an on-shell Bethe vector. In this way various integral representations were obtained for correlation functions
of the $XXX$ and $XXZ$ spin-$1/2$ chains and of the model of one-dimensional bosons
\cite{KitMT00,KitMST02,KitKMST09b,GohKS04,GohKS05,SeeBGK07}.
Determinant formulas were also used for numerical analysis of correlation functions
\cite{CauHM05,PerSCHMWA06,PerSCHMWA07,CauCS07}.

A wide class of quantum integrable models is associated with higher rank algebras $\frak{gl}_N$.
An algebraic Bethe ansatz for these type models is called hierarchical (or nested) and was
introduced in \cite{KulR83} (see also \cite{VarT95,VarT07,BelR08}). The first
result concerning the scalar products in the models with $SU(3)$-invariant
$R$-matrix was obtained by Reshetikhin in \cite{Res86}. There, an analog of Izergin--Korepin
formula  for the scalar product of generic off-shell Bethe vectors  and a determinant representation
for the norm of the on-shell vectors were found. Recently various particular cases of scalar products
were studied in \cite{BelPR,Whe12,PozOK12,EGSV}.

In this paper we study a particular case of the scalar products in a generalized model with the $SU(3)$-invariant $R$-matrix
 \be{R-mat}
 R(x,y)=\mathbf{I}+g(x,y)\mathbf{P}, \qquad g(x,y)=\frac{c}{x-y},
 \ee
where $\mathbf{I}$ is the identity matrix, $\mathbf{P}$ is the permutation matrix, $c$ is a constant.
Keeping in mind possible generalization of our results to models with $q$-deformed $SU(3)$-symmetry
we do not stress that the function $g(x,y)$ depends on the difference $x-y$.

The result obtained in this paper is a determinant
representation for the product of an on-shell Bethe vector and an eigenvector of the twisted
transfer-matrix (twisted on-shell vector). The notion of a twisted transfer-matrix was found to be very useful for evaluation
of certain correlation functions in $\frak{gl}_2$-based integrable models \cite{KitMST05}. One can use this object
in the case of higher rank algebras as well. The twisted monodromy matrix is introduced as follows. The monodromy matrix $T(w)$
satisfies the algebra
\be{RTT}
R_{12}(w_1,w_2)T_1(w_1)T_2(w_2)=T_2(w_2)T_1(w_1)R_{12}(w_1,w_2).
\ee
Equation \eqref{RTT} holds in the tensor product $V_1\otimes V_2\otimes\mathcal{H}$,
where $V_k\sim\mathbb{C}^3$ are auxiliary linear spaces, and $\mathcal{H}$ is the Hilbert space of the Hamiltonian for the model under consideration.
The $R$-matrix acts non-trivially in $V_1\otimes V_2$, the matrices $T_k$ act non-trivially in
$V_k\otimes \mathcal{H}$. Let  $\rho$ be a matrix such that its tensor square commutes with the $R$-matrix:
$[\rho_1\rho_2,R_{12}]=0$. Define a  twisted monodromy matrix as $\tilde T=\rho T$.
Then it is easy to see that
\be{tRTT}
R_{12}(w_1,w_2)\tilde T_1(w_1)\tilde T_2(w_2)=\tilde T_2(w_2)\tilde T_1(w_1)R_{12}(w_1,w_2),
\ee
that is, the twisted monodromy matrix enjoys the same algebra as the original one. Thus, the
eigenvectors of the twisted transfer-matrix $\tr \tilde T(w)$ can be found in the framework of the standard scheme.
Below we consider a special case of the matrix $\rho=\diag(1,\kappa,1)$,
where $\kappa$ is a complex number (twist parameter). Therefore we denote the twisted
monodromy matrix by $T_\kappa(w)$.

The article is organized as follows. In section~\ref{S-N} we introduce the model under consideration and describe
the notations used in the paper. Section~\ref{S-MR} contains the determinant representation for the scalar
product of twisted and standard on-shell Bethe vectors. This is the main result of the paper.
In section~\ref{S-XXX} we show how our result
can be applied to the calculation of form factors of some local operators in the $SU(3)$-invariant XXX chain.
In section~\ref{S-ST} we consider several particular cases of the obtained determinant representation. In particular,
we reproduce the formula for the norm of on-shell Bethe vectors \cite{Res86}.
Finally, in  section~\ref{S-CSP} we give the derivation of the determinant representation for the scalar product.
Appendix \ref{A-IHC} presents some properties of the domain wall partition function (DWPF), and
Appendices \ref{Pmain-ident}--\ref{Wonderful} gather  several auxiliary
Lemmas.

\section{Notations\label{S-N}}

First of all we give a list of notations and conventions used in the paper.

Apart from the function $g(x,y)$ we also introduce a function $f(x,y)$ as
\be{univ-not}
 f(x,y)=\frac{x-y+c}{x-y}.
\ee
Clearly that in our case $f(x,y)=1+g(x,y)$; however the $q$-analogues of these functions do not satisfy this property.
Two other auxiliary functions  will be also used
\be{desand}
h(x,y)=\frac{f(x,y)}{g(x,y)}=\frac{x-y+c}{c},\qquad  t(x,y)=\frac{g(x,y)}{h(x,y)}=\frac{c^2}{(x-y)(x-y+c)}.
\ee
The following obvious properties of the functions introduced above are useful
 \be{propert}
 g(x,y)=-g(y,x),\quad h(x-c,y)=g^{-1}(x,y),\quad  f(x-c,y)=f^{-1}(y,x),\quad  t(x-c,y)=t(y,x).
 \ee

Before giving a description of the Bethe vectors we formulate a convention on the notations.
We always denote sets of variables by bar: $\bar w$, $\blac$, $\bmub$ etc.
Individual elements
of the sets are denoted by subscripts: $w_j$, $\lab_k$ etc. As a rule, the number of elements in the
sets is not shown explicitly in the equations; however, we give these cardinalities in
special comments after the formulas.
 Subsets of variables are denoted by roman indices: $\blac_{\so}$, $\bmub_{\rm iv}$, $\bar w_{\st}$ etc.
We assume
that the elements in every subset of variables are ordered in such a way that the sequence of
their subscripts is strictly increasing. We call this ordering  natural order.

In order to avoid too cumbersome formulas we use shorthand notations for products of scalar
functions. Namely, if functions $g$, $f$, $h$, $t$, as well as $r_1$ and $r_3$ (see \eqref{ratios}), depend
on sets of variables, this means that one should take the product over the corresponding set.
For example,
 \be{SH-prod}
 r_1(\blac_{\st})=\prod_{\lac_j\in\blac_{\st}} r_1(\lac_j);\quad
 g(\mub_k, \bar w)= \prod_{w_j\in\bar w} g(\mub_k, w_j);\quad
 f(\blab_{\st},\blab_{\so})=\prod_{\lab_j\in\blab_{\st}}\prod_{\lab_k\in\blab_{\so}} f(\lab_j,\lab_k).
 \ee

Now we pass to the description of Bethe vectors.
A generic Bethe vector is denoted by $|\bla,\bmu\rangle$.
It is parameterized by two sets of
complex parameters $\bla=\la_1,\dots,\la_a$ and $\bmu=\muu_1,\dots,\muu_b$ with $a,b=0,1,\dots$.
Dual Bethe vectors are denoted by $\langle\bmu,\bla|$. They also depend on two sets of
complex parameters $\bla=\la_1,\dots,\la_a$ and $\bmu=\muu_1,\dots,\muu_b$. The state with
$\bla=\bmu=\emptyset$ is called a pseudovacuum vector $|0\rangle$. Similarly the dual state
with $\bla=\bmu=\emptyset$ is called a dual pseudovacuum vector $\langle0|$. These vectors
are annihilated by the operators $T_{jk}(w)$, where $j>k$ for  $|0\rangle$ and $j<k$ for $\langle0|$.
At the same time both vectors are eigenvectors for the diagonal entries of the monodromy matrix
 \be{Tjj}
 T_{jj}(w)|0\rangle=\as_j(w)|0\rangle, \qquad   \langle0|T_{jj}(w)=\as_j(w)\langle0|,
 \ee
where $\as_j(w)$ are some scalar functions. In the framework of the generalized model, $\as_j(w)$ remain free functional parameters. Actually, it is always possible to normalize
the monodromy matrix $T(w)\to \as_2^{-1}(w)T(w)$ so as to deal only with the ratios
 \be{ratios}
 r_1(w)=\frac{\as_1(w)}{\as_2(w)}, \qquad  r_3(w)=\frac{\as_3(w)}{\as_2(w)}.
 \ee

On-shell Bethe vectors are eigenvectors of the transfer matrix, while twisted on-shell Bethe vectors are eigenvectors of the twisted transfer matrix. We will consider the scalar product of on-shell vector and dual twisted on-shell  vector. Then
\be{Left-act}
\tr T(w)|\blab,\bmub\rangle = \tau(w|\blab,\bmub)\,|\blab,\bmub\rangle,\qquad
 \langle\blac,\bmuc|\tr T_\kappa(w) = \tau_\kappa(w|\blac,\bmuc)\,\langle\blac,\bmuc|,
\ee
where
\be{tau-def}
\begin{array}{l}
\tau(w)\equiv\tau(w|\blab,\bmub)=r_1(w)f(\blab,w)+f(w,\blab)f(\bmub,w)+r_3(w)f(w,\bmub),\\
\tau_\kappa(w)\equiv\tau_\kappa(w|\blac,\bmuc)=r_1(w)f(\blac,w)+\kappa f(w,\blac)f(\bmuc,w)+r_3(w)f(w,\bmuc).
\end{array}
\ee
Hereby  the sets $\blab$ and $\bmub$ (respectively $\blac$ and $\bmuc$) satisfy the system of nested Bethe ansatz
equations (respectively the twisted system of nested Bethe ansatz
equations) \cite{KulR83}. We give these systems in a slightly unusual form
\be{AEigenS-1}
r_1(\blab_{\so})=\frac{f(\blab_{\so},\blab_{\st})}{f(\blab_{\st},\blab_{\so})}f(\bmub,\blab_{\so}),
\ee
\be{AEigenS-2}
r_3(\bmub_{\so})=\frac{f(\bmub_{\st},\bmub_{\so})}{f(\bmub_{\so},\bmub_{\st})}f(\bmub_{\so},\blab).
\ee
These equations should hold for arbitrary partitions of the sets $\blab$ and $\bmub$ into subsets
$\{\blab_{\so},\;\blab_{\st}\}$ and $\{\bmub_{\so},\;\bmub_{\st}\}$ respectively. Obviously, it is enough
to demand that the system  \eqref{AEigenS-1}, \eqref{AEigenS-2} is valid for the
particular case, when the sets $\blab_{\so}$ and $\bmub_{\so}$ consist of only one element. Then it  turns into the standard system of Bethe equations.

The twisted equations have the similar form
\be{ATEigenS-1}
r_1(\blac_{\so})=\kappa^{k_{\so}}\frac{f(\blac_{\so},\blac_{\st})}{f(\blac_{\st},\blac_{\so})}f(\bmuc,\blac_{\so}),
\ee
\be{ATEigenS-2}
r_3(\bmuc_{\so})=\kappa^{n_{\so}}\frac{f(\bmuc_{\st},\bmuc_{\so})}{f(\bmuc_{\so},\bmuc_{\st})}f(\bmuc_{\so},\blac),
\ee
where $k_{\so}=\#\blac_{\so}$ and $n_{\so}=\#\bmuc_{\so}$.

The scalar products of Bethe vectors are defined as
 \be{SP-deFF}
 \mathcal{S}_{a,b}\equiv\mathcal{S}_{a,b}(\blac,\blab|\bmuc,\bmub)=
 \frac{\langle\blac,\bmuc|\blab,\bmub\rangle}{\as_2(\blac)\as_2(\blab)\as_2(\bmuc)\as_2(\bmub)}.
 \ee
If $|\blab,\bmub\rangle$ and $\langle\blac,\bmuc|$ are generic off-shell Bethe vectors, then the parameters $\blac$, $\blab$, $\bmuc$, and $\bmub$ are arbitrary complex numbers. We will
consider the particular case of $\mathcal{S}_{a,b}$, when $\blab$ and $\bmub$ satisfy the system
\eqref{AEigenS-1}, \eqref{AEigenS-2}, while $\blac$ and $\bmuc$ satisfy the system
\eqref{ATEigenS-1}, \eqref{ATEigenS-2}.

To conclude this section we introduce the partition function of the six-vertex model with domain wall boundary conditions (DWPF) \cite{Kor82,Ize87}. This is one of the central object in the study of scalar products of $\frak{gl}_2$-based models.   It also plays an important role in the case of
$SU(3)$ invariant mosels. We denote the DWPF by
$K_n(\bar x|\bar y)$. It depends on two sets of variables $\bar x$ and $\bar y$, the subscript shows that
$\#\bar x=\#\bar y=n$. The function $K_n$ has the following determinant representation:
\begin{equation}\label{K-def}
K_n(\bar x|\bar y)
=\Delta'_n(\bar x)\Delta_n(\bar y)h(\bar x,\bar y)
\det_n t(x_j,y_k),
\end{equation}
where $\Delta'_n(\bar x)$ and $\Delta_n(\bar y)$ are
\be{def-Del}
\Delta'_n(\bar x)
=\prod_{j>k}^n g(x_j,x_k),\qquad {\Delta}_n(\bar y)=\prod_{j<k}^n g(y_j,y_k).
\ee
It is easy to see that $K_n$ is symmetric over $\bar x$ and symmetric over $\bar y$; however,  $K_n(\bar x|\bar y)\ne
 K_n(\bar y|\bar x)$. Below we consider
$K_n$ depending on combinations of sets of different variables, for example $K_{n}(\bar\xi|\bar \alpha,\bar \beta+c)$.
Due to the symmetry properties of the DWPF $K_{n}(\bar\xi|\bar \alpha,\bar \beta+c)=K_{n}(\bar\xi|\bar \beta+c,\bar \alpha)$.

\section{Results\label{S-MR}}

The scalar product of on-shell and twisted on-shell Bethe vectors has the following determinant
representation:
 \begin{equation}\label{R-fin1}
 \mathcal{S}_{a,b}= f(\bmuc,\blac)f(\bmub,\blab)t(\bmuc,\blab)
 \Delta'_a(\blac)\Delta_a(\blab)\Delta'_b(\bmuc)\Delta_b(\bmub)
 \det_{a+b}\mathcal{N},
 \end{equation}
where $\mathcal{N}$ is a block-matrix of the size $(a+b)\times(a+b)$,
\begin{equation}\label{R-fin2}
 \mathcal{N}=\left(\begin{array}{cc}
 {\dis \mathcal{N}^{(\la)}(\lac_j,\lab_k)}&
 {\dis \mathcal{N}^{(\la)}(\lac_j,\muc_k)}\num
 {\dis \mathcal{N}^{(\muu)}(\mub_j,\lab_k)}&
 {\dis \mathcal{N}^{(\muu)}(\mub_j,\muc_k)}
 \end{array}\right).
 \end{equation}
%
The diagonal block $\mathcal{N}^{(\la)}(\lac_j,\lab_k)$ is an $a\times a$ matrix with  entries
 \be{R-P11}
 \mathcal{N}^{(\la)}(\lac_j,\lab_k)=h(\bmuc,\lab_k)h(\lab_k,\blac)
 \left[\kappa t(\lab_k,\lac_j)+t(\lac_j,\lab_k)\frac{f(\bmub,\lab_k)}{f(\bmuc,\lab_k)}
 \frac{h(\blac,\lab_k)h(\lab_k,\blab)}{h(\lab_k,\blac)h(\blab,\lab_k)} \right].
 \ee
The entries of a $b\times b$ block $\mathcal{N}^{(\muu)}(\mub_j,\muc_k)$ are
 \be{R-P22}
 \mathcal{N}^{(\muu)}(\mub_j,\muc_k)=h(\muc_k,\blab)h(\bmub,\muc_k)\left[t(\mub_j,\muc_k)+
 \kappa t(\muc_k,\mub_j)\frac{f(\muc_k,\blac)}{f(\muc_k,\blab)}
 \frac{h(\bmuc,\muc_k )h(\muc_k,\bmub)}{h(\muc_k,\bmuc)h(\bmub,\muc_k )}
    \right].
 \ee
The off-diagonal blocks have asimpler form
 \be{R-P12}
 \mathcal{N}^{(\la)}(\lac_j,\muc_k)=\kappa
  t(\muc_k,\lac_j)h(\muc_k,\blac)h(\bmuc,\muc_k),
 \ee
 and
 \be{R-P21}
 \mathcal{N}^{(\muu)}(\mub_j,\lab_k)=
 t(\mub_j,\lab_k)h(\bmub,\lab_k )h(\lab_k,\blab).
 \ee
 %


Observe that the entries of the matrix $\mathcal{N}$ can be written in a more compact form in terms
of Jacobians of the transfer-matrix and the twisted transfer-matrix eigenvalues (similarly to \cite{KitMT99})
 \be{Upper-tau}
 \mathcal{N}^{(\la)}(\lac_j,w_k)= \lim_{w\to w_k}\Big(
 cg^{-1}(w,\blac)g^{-1}(\bmuc,w)
 \frac{\partial \tau_\kappa(w)}{\partial\lac_j}\Big),
 \qquad w_k=\lab_k \quad\text{or}\quad w_k= \muc_k,
 \ee
and
 \be{Down-tau}
 \mathcal{N}^{(\muu)}(\mub_j,w_k)= \lim_{w\to w_k}\Big(-cg^{-1}(\bmub,w )g^{-1}(w,\blab)
 \frac{\partial \tau(w)}{\partial\mub_j}\Big),\qquad w_k=\lab_k \quad\text{or}\quad w_k= \muc_k.
 \ee
In the calculation of partial derivatives in these formulas, one has to consider $r_1(w)$ and $r_3(w)$ as arbitrary functions. Then, to compute the limit, one
should express $r_1(\lab_k)$ and
$r_3(\muc_k)$ via \eqref{AEigenS-1}, \eqref{ATEigenS-2}. Altogether, they turn into \eqref{R-P11}--\eqref{R-P21}.


\section{Form factors in $SU(3)$ invariant XXX chain\label{S-XXX}}

In this section we consider an application of our result to the calculation of form factors
of local operators of the $SU(3)$-invariant XXX chain. This model is a
particular case of the generalized model considered in this paper.

The monodromy matrix of the periodic XXX chain with $N$ sites is given as a product of $R$-matrices
 \be{Mon-Mat}
 T(w)=R_{0N}(w,0)\cdots R_{02}(w,0)R_{01}(w,0).
 \ee
Here $R_{0m}$ acts in $V_0\otimes V_m$, where
$V_0\sim\mathbb{C}^3$ is the the auxiliary space and $V_m\sim\mathbb{C}^3$ is the local quantum space  associated with the $m$th site of the chain. Thus, $T(w)$ is a
$3\times 3$ matrix in the space $V_0$, whose entries are operators in $\otimes_{m=1}^N V_m$.

The local operators acting in the quantum space $V_m$  are elementary units
 \be{El-Un}
 E^{\epsilon,\epsilon'}_m, \qquad \epsilon,\epsilon'=1,2,3, \qquad
 \left(E^{\epsilon,\epsilon'}\right)_{jk}=\delta_{j\epsilon}\delta_{k\epsilon'}.
 \ee
The inverse scattering problem for this model was solved in \cite{MaiT00}.
Using this solution one can express the  operators $E^{\epsilon,\epsilon'}_m$  in terms of the entries of the monodromy
matrix\footnote[1]{Due to \eqref{Mon-Mat} $T(w)$ is singular at $w=0$.  One should understand the equation \eqref{gen-sol-T}
as a limit $w\to 0$}
 \be{gen-sol-T}
 E^{\epsilon,\epsilon'}_m =(\tr T(0))^{m-1}  T_{\epsilon',\epsilon}(0)(\tr T(0))^{-m}.
 \ee

Suppose that we want to compute correlation functions involving the operators $E^{2,2}_m$. For this
we consider a form factor of the operator $E^{2,2}_m$
 \be{ffp}
F_m^{2,2}=\langle\tilde\psi| E^{2,2}_m|\psi\rangle.
\ee
Here $|\psi\rangle$ and $\langle\tilde\psi|$ are some on-shell Bethe vectors. Let $Q_m=\sum\limits_{k=1}^m E^{2,2}_k$. Obviously
 \be{ffpQ}
E^{2,2}_m=\left.\frac{d}{d\beta}(\eee^{\beta Q_m}-\eee^{\beta Q_{m-1}})\right|_{\beta=0}.
\ee
Due to  $(E^{2,2}_k)^n=E^{2,2}_k$ for $n\ge 1$ we have
 \be{bQ}
\eee^{\beta Q_m}=\prod_{k=1}^m \eee^{\beta E^{2,2}_k}=\prod_{k=1}^m\left(1+E^{2,2}_k (\eee^{\beta }-1)\right)=
 \prod_{k=1}^m\left(E^{1,1}_k+\eee^{\beta }E^{2,2}_k +E^{3,3}_k\right).
 \ee
Using the solution of the quantum inverse scattering problem \eqref{gen-sol-T} we obtain
 \be{p-T}
 E^{1,1}_k+\eee^{\beta }E^{2,2}_k +E^{3,3}_k=(\tr T(0))^{k-1} \tr T_\kappa(0)(\tr T(0))^{-k},
 \ee
where we have set $\kappa=\eee^\beta$.  Then, due to \eqref{bQ}, we find
 \be{bQ1}
 \eee^{\beta Q_m}= (\tr T_\kappa(0))^m(\tr T(0))^{-m}.
 \ee

Let now $\langle\tilde\psi_\kappa|$ be a  twisted on-shell Bethe vector, such that
$\langle\tilde\psi_\kappa|\to\langle\tilde\psi|$ at $\kappa\to 1$ (that is, at $\beta\to 0$). Consider
\be{QTb}
F_m^\beta=\langle\tilde\psi_\kappa| \eee^{\beta Q_m}|\psi\rangle.
\ee
Using \eqref{bQ1} we immediately obtain
\be{QTb-a}
F_m^\beta=\langle\tilde\psi_\kappa| (\tr T_\kappa(0))^m(\tr T(0))^{-m}|\psi\rangle=\frac{\tilde\tau_\kappa^m(0)}{\tau^m(0)}\langle\tilde\psi_\kappa|\psi\rangle.
\ee
Here $\tau(0)$ is the eigenvalue of $\tr T(0)$ on $|\psi\rangle$ and $\tilde\tau_\kappa(0)$ is the eigenvalue
of $\tr T_\kappa(0)$ on $\langle\tilde\psi_\kappa|$. Now we simply take derivative of $F_m^\beta$
over $\beta$ at $\beta=0$. Differentiating \eqref{QTb} we find
\be{d-QTb}
\left.\frac{dF_m^\beta}{d\beta}\right|_{\beta=0}=\langle\tilde\psi| Q_m|\psi\rangle
+\left.\frac{d}{d\beta}\langle\tilde\psi_\kappa|\psi\rangle\right|_{\beta=0}.
\ee
On the other hand, taking the derivative of \eqref{QTb-a} we obtain
\be{d-QTb1}
\left.\frac{dF_m^\beta}{d\beta}\right|_{\beta=0}=
\left.\frac{d}{d\beta}\frac{\tilde\tau_\kappa^m(0)}{\tau^m(0)}\right|_{\beta=0}\cdot
\langle\psi|\psi\rangle\,\delta_{\tilde\psi,\psi}
+\frac{\tilde\tau^m(0)}{\tau^m(0)}\cdot\left.\frac{d}{d\beta}\langle\tilde\psi_\kappa|\psi\rangle\right|_{\beta=0},
\ee
where $\delta_{\tilde\psi,\psi}=1$ if $\langle\tilde\psi|=|\psi\rangle^\dagger$ and $\delta_{\tilde\psi,\psi}=0$
otherwise. $\langle\psi|\psi\rangle$ is the norm of the vector $|\psi\rangle$, as computed in section \ref{S-Norm}. Comparing \eqref{d-QTb} and \eqref{d-QTb1} we arrive at
\be{QQ}
\langle\tilde\psi| Q_m|\psi\rangle=
\left.\frac{d}{d\beta}\frac{\tilde\tau_\kappa^m(0)}{\tau^m(0)}\right|_{\beta=0}\cdot
\langle\psi|\psi\rangle\,\delta_{\tilde\psi,\psi}
+\left(\frac{\tilde\tau^m(0)}{\tau^m(0)}-1\right)\cdot\left.\frac{d}{d\beta}\langle\tilde\psi_\kappa|\psi\rangle\right|_{\beta=0}.
\ee
It remains to use that $E^{2,2}_m =Q_m-Q_{m-1}$ and we obtain the representation of the form
factor $\langle\tilde\psi| E^{2,2}_m|\psi\rangle$ in terms of the scalar product
$\langle\tilde\psi_\kappa|\psi\rangle$ between twisted and standard on-shell Bethe vectors,
\be{QQ2}
\langle\tilde\psi| E^{2,2}_m|\psi\rangle=
\frac{d}{d\kappa}\left[\Big(\frac{\tilde\tau_\kappa^m(0)}{\tau^m(0)}-
\frac{\tilde\tau_\kappa^{m-1}(0)}{\tau^{m-1}(0)}\Big)\cdot
\langle\tilde\psi_\kappa|\psi\rangle\right]_{\kappa=1}.
\ee

\section{Several tests \label{S-ST}}

In this section we make several tests of our result. In particular, at $\kappa=1$ we
deal with the scalar product of two on-shell Bethe vectors. Hence, we should obtain either zero
for $\langle\blac,\bmuc|\ne|\blab,\bmub\rangle^\dagger$, or the expression for the
norm \cite{Res86} if $\langle\blac,\bmuc|=|\blab,\bmub\rangle^\dagger$.

\subsection{Norm of eigenvector\label{S-Norm}}

To approach the case of the norm we should set $\blac=\blab$, $\bmuc=\bmub$, and $\kappa=1$. There is no problem
to take this limit in the off-diagonal blocks \eqref{R-P12} and \eqref{R-P21}. In the diagonal blocks the limit
is a bit more complicated. It is more convenient to take this limit in the equations \eqref{Upper-tau} and \eqref{Down-tau}.
Taking the partial derivatives we obtain for the diagonal blocks
 \be{P11}
 \mathcal{N}^{(\la)}(\lac_j,\lab_k)=h(\bmuc,\lab_k)\Bigl[(-1)^{a+1}\frac{r_1(\lab_k)}{f(\bmuc,\lab_k)}
 t(\lac_j,\lab_k)h(\blac,\lab_k) +\kappa
  t(\lab_k,\lac_j)h(\lab_k,\blac)\Bigr],
 \ee
and
 \be{P22}
 \mathcal{N}^{(\muu)}(\mub_j,\muc_k)=h(\muc_k,\blab)\Bigl[(-1)^{b+1}\frac{r_3(\muc_k)}{f(\muc_k,\blab)}
  t(\muc_k,\mub_j)h(\muc_k,\bmub) +
 t(\mub_j,\muc_k)h(\bmub,\muc_k )\Bigr].
 \ee
Consider, for example, the block \eqref{P11}. Let $\lab_k=\lac_k+\epsilon$ and
$X_k^{(1)}=r'_1(\lac_k)/r_1(\lac_k)$. Then
up to $\epsilon^2$ terms we have
 \be{dev-r1}
 r_1(\lab_k)=r_1(\lac_k)(1+\epsilon X_k^{(1)})=(-1)^{a+1}\kappa\,(1+\epsilon X_k^{(1)})
\frac{h(\blac,\lac_k)}{h(\lac_k,\blac)}f(\bmuc,\lac_k),
\ee
where we have used \eqref{ATEigenS-1}, \eqref{desand} and \eqref{propert}. Substituting this into \eqref{P11} and setting $\lac_k\equiv\la_k$ we obtain
for diagonal entries of the matrix $\mathcal{N}^{(\la)}(\lac_k,\lab_k)$
 \begin{multline}\label{P11-n1}
 \mathcal{N}^{(\la)}(\la_k,\la_k+\epsilon)=\kappa h(\bmuc,\la_k)h(\la_k,\bla)\num
 \times\left[t(\la_k+\epsilon,\la_k) +t(\la_k,\la_k+\epsilon) (1+\epsilon X_k^{(1)})
   \frac{f(\bmuc,\la_k)}{f(\bmuc,\la_k+\epsilon)}
 \frac{h(\bla,\la_k+\epsilon)h(\la_k,\bla)}{h(\bla,\la_k)h(\la_k+\epsilon,\bla)}
  \right].
 \end{multline}
Now we can safely proceed to the limit $\epsilon=0$. This gives us
 \begin{equation}\label{P11-n2}
 \mathcal{N}^{(\la)}(\la_k,\la_k)=\kappa h(\bmuc,\la_k)h(\la_k,\bla)
 \left[-cX_k^{(1)}-2-\sum_{\ell=1}^a\frac{2c^2}{\la_{k\ell}^2-c^2}+\sum_{m=1}^b
 t(\muc_m,\la_k)  \right],
 \end{equation}
where $\la_{k\ell}=\la_k-\la_\ell$.

In the off-diagonal entries of the matrix $\mathcal{N}^{(\la)}(\lac_j,\lab_k)$ we can simply
set $\lac_j=\la_j$ and $\lab_k=\la_k$ and use \eqref{ATEigenS-1}.  Altogether, the diagonal block \eqref{P11} takes the form
 \begin{multline}\label{P11-n3}
 \mathcal{N}^{(\la)}(\la_j,\la_k)= h(\bmu,\la_k)h(\la_k,\bla)\num
 \times\left[\delta_{jk}\left(-cX_k^{(1)}-\sum_{\ell=1}^a\frac{2c^2}{\la_{k\ell}^2-c^2}+\sum_{m=1}^b
 t(\muu_m,\la_k)\right)  +\frac{2c^2}{\la_{jk}^2-c^2}\right].
 \end{multline}
Here we have already set $\kappa=1$ and $\bmuc=\bmub\equiv\bmu$.


Note that it would be much more sophisticated
to take the limit $\blac=\blab$ in expression \eqref{R-P11}. Then we
should consider the parameters $\blac$ as functions of $\kappa$: $\lac_j\equiv\lac_j(\kappa)$ and
take the limit $\kappa\to1$. It is clear that after taking this limit we obtain an expression containing
derivatives $d\lac_j/d\kappa$ at $\kappa=1$, and one should prove that all these derivatives can be
combined into $X_k^{(1)}$. Certainly this method is more cumbersome than the technics described above.

The limit $\bmuc=\bmub\equiv\bmu$ in the matrix $\mathcal{N}^{(\muu)}(\mub_j,\muc_k)$
can be taken in a similar way.  After
simple algebra we obtain for the norm of eigenvector
 \begin{equation}\label{Norm-fin}
 \mathcal{S}_{a,b}(\bla,\bla|\bmu,\bmu)=f^3(\bmu,\bla)\prod_{j,k=1\atop{j\ne k}}^af(\la_j,\la_k)
 \prod_{j,k=1\atop{j\ne k}}^b f(\muu_j,\muu_k)\cdot
 \det_{a+b}\left(\begin{array}{cc}
 {\dis \mathcal{N}_{11}(\la_j,\la_k)}&
t(\muu_k,\la_j)\num
 t(\la_k,\muu_j)&
 {\dis \mathcal{N}_{22}(\muu_j,\muu_k)}
 \end{array}\right).
 \end{equation}
Here the diagonal blocks are
 \begin{equation}\label{P11-n4}
\mathcal{N}_{11}(\la_j,\la_k)=
 \delta_{jk}\left(-cX_k^{(1)}-\sum_{\ell=1}^a\frac{2c^2}{\la_{k\ell}^2-c^2}+\sum_{m=1}^b
 t(\muu_m,\la_k)\right)  +\frac{2c^2}{\la_{jk}^2-c^2},
 \end{equation}
 \begin{equation}\label{P22-n}
\mathcal{N}_{22}(\muu_j,\muu_k)=\delta_{jk}\left(cX_k^{(3)}-\sum_{m=1}^b\frac{2c^2}{\muu_{km}^2-c^2}+\sum_{\ell=1}^a
 t(\muu_k,\la_\ell)\right)  +\frac{2c^2}{\muu_{jk}^2-c^2},
 \end{equation}
where $X_k^{(3)}=r'_3(\muu_k)/r_3(\muu_k)$ and $\muu_{jk}=\muu_j-\muu_k$.
%
 %
This result coincides with the one of \cite{Res86} up to notations.

\subsection{Scalar product of two different on-shell vectors\label{SS-ZEV}}

If $\blac\ne\blab$ or $\bmuc\ne\bmub$ at $\kappa=1$, then the result obtained describes the scalar product of two
different eigenvectors of the transfer-matrix. Hence, we should have $\mathcal{S}_{a,b}=0$ in this case. Below we prove
that $\mathcal{S}_{a,b}$ does vanish at $\kappa=1$ except for the case of the norm. For this
we construct an eigenvector of the block-matrix $\mathcal{N}$ \eqref{R-fin1} having zero
eigenvalue at $\kappa=1$. This means that the determinant in \eqref{R-fin1} vanishes at $\kappa=1$.

The zero eigenvector has the following components
\be{def-Omega}
\begin{array}{l}
{\dis \Omega_k=\prod\limits_{\ell=1}^a(\lac_k-\lab_\ell)
\prod\limits_{\ell=1\atop{\ell\ne k}}^a(\lac_k-\lac_\ell)^{-1},\qquad k=1,\dots,a,}\num
{\dis \Omega_{a+k}=\prod\limits_{m=1}^b(\mub_k-\muc_m)
\prod\limits_{m=1\atop{m\ne k}}^b(\mub_k-\mub_m)^{-1},\qquad k=1,\dots,b.}
\end{array}
\ee
We should prove that
\be{act-Omega}
\begin{array}{l}
{\dis \sum_{j=1}^a\Omega_j
\mathcal{N}^{(\la)}(\lac_j,\lab_k)+\sum_{j=1}^b
 \Omega_{a+j}\mathcal{N}^{(\muu)}(\mub_j,\lab_k)=0,}\num
 {\dis \sum_{j=1}^a\Omega_j\mathcal{N}^{(\la)}(\lac_j,\muc_k)+
 \sum_{j=1}^b\Omega_{a+j}\mathcal{N}^{(\muu)}(\mub_j,\muc_k)=0.}
\end{array}
\ee

Equations \eqref{act-Omega} can be checked straightforwardly. Let us  compute,
for instance, the action of the left blocks \eqref{R-P11} and \eqref{R-P21} on the vector $\Omega$.
Assume that $\lab_j\ne\lac_k$, $\forall j,k=1,\dots a$ and $\mub_j\ne\muc_k$, $\forall j,k=1,\dots b$.
Then in order to calculate the sum
 \be{sim-sum}
 \sum_{j=1}^a\Omega_j
\mathcal{N}^{(\la)}(\lac_j,\lab_k),
\ee
we introduce
 \be{Hpm}
 H^\pm_k=\sum_{j=1}^a\frac{c^2}{(\lac_j-\lab_k)(\lac_j-\lab_k\pm c)}
 \prod\limits_{\ell=1}^a(\lac_j-\lab_\ell)
\prod\limits_{\ell=1\atop{\ell\ne j}}^a(\lac_j-\lac_\ell)^{-1}.
\ee
The sums $H^\pm_k$ can be computed by means of an auxiliary integral
\be{Integ}
I=\frac1{2\pi i}\oint\limits_{|z|=R\to\infty}
\frac{c^2\,dz}{(z-\lab_k)(z-\lab_k\pm c)}
\prod\limits_{\ell=1}^a \frac{z-\lab_\ell}{z-\lac_\ell}.
\ee
The integral is taken over the contour $|z|=R$ and we consider the limit $R\to\infty$. Then $I=0$, because
the integrand behaves as $1/z^2$ at $z\to\infty$. On the other hand the same integral is equal to the sum
of the residues within the integration contour. Obviously the sum of the residues at $z=\lac_\ell$ gives
$H^\pm_k$. There is also one additional pole at $z-\lab_k\pm c=0$. Then we have
 \be{I-res}
 I=0=H^\pm_k \mp c \prod_{\ell=1}^a\frac{\lab_k-\lab_\ell\mp c}{\lab_k-\lac_\ell\mp c}.
 \ee
From this we find
 \be{act-1}
 \begin{array}{l}
{\dis \sum_{j=1}^a\Omega_j t(\lac_j,\lab_k)=c\frac{h(\blab,\lab_k)}{h(\blac,\lab_k)},}\num
{\dis \sum_{j=1}^a\Omega_j t(\lab_k,\lac_j)=-c\frac{h(\lab_k,\blab)}{h(\lab_k,\blac)}.}
\end{array}
\ee
Using these results we immediately obtain
\be{act-P11}
\sum_{j=1}^a\Omega_j \mathcal{N}^{(\la)}(\lac_j,\lab_k)=
c\, h(\bmuc,\lab_k)h(\lab_k,\blab)\left(\frac{f(\bmub,\lab_k)}{f(\bmuc,\lab_k)}-1\right).
\ee

The action of the block $\mathcal{N}^{(\muu)}(\mub_j,\lab_k)$ on $\Omega_{a+j}$ can be calculated in
the similar way. Using an auxiliary contour integral
\be{Integ-1}
I=\frac1{2\pi i}\oint\limits_{|z|=R\to\infty}
\frac{c^2\,dz}{(z-\lab_k)(z-\lab_k+ c)}
\prod\limits_{m=1}^b \frac{z-\muc_m}{z-\mub_m},
 \ee
we find that
 \be{act-2}
\sum_{j=1}^a\Omega_{a+j} t(\mub_j,\lab_k)=c\frac{h(\bmuc,\lab_k)}{h(\bmub,\lab_k)}
\left(1-\frac{f(\bmub,\lab_k)}{f(\bmuc,\lab_k)}\right).
\ee
This implies
\be{act-P21}
\sum_{j=1}^b\Omega_{a+j} \mathcal{N}^{(\muu)}(\mub_j,\lab_k)=
c\, h(\bmuc,\lab_k)h(\lab_k,\blab)\left(1-\frac{f(\bmub,\lab_k)}{f(\bmuc,\lab_k)}\right),
\ee
and hence, the first equation \eqref{act-Omega} is proved. The proof of the second equation \eqref{act-Omega}
is similar.

The proof given above should be slightly modified in the case when some parameters $\lac_j$ and $\lab_j$
(or $\muc_j$ and $\mub_j$) coincide. Consider without loss of generality the case $\lac_j=\lab_j$ for
$j=1,\dots,n$ and $\lac_j\ne\lab_j$ for $j=n+1,\dots,a$. We also assume that $\muc_j\ne\mub_j$ for
$j=1,\dots,b$. Then first of all we should take the limit $\lac_j\to\lab_j$ in the first $n$ rows of the block
$\mathcal{N}^{(\la)}(\lac_j,\lab_k)$,
just like we did in section~\ref{S-Norm}. As a result the explicit expressions for the corresponding entries change; in particular, they depend on the logarithmic derivatives $X^{(1)}_k$. However, at the same time, the first $n$ elements of the
vector $\Omega$ vanish:  $\Omega_{j}=0$, for $j=1,\dots,n$. Therefore, independently of the explicit form
of the entries in the first $n$ rows of the block $\mathcal{N}^{(\la)}(\lac_j,\lab_k)$, their action on the vector $\Omega$ gives zero.
The action of the remaining part of the block $\mathcal{N}^{(\la)}(\lac_j,\lab_k)$ can be calculated exactly in the same way as we did before. One can easily check that the same vector $\Omega$ remains the zero eigenvector of the matrix $\mathcal{N}$.

The only case where the proof fails is the case $\blab=\blac$ and $\bmub=\bmuc$, that is, the case of the norm. In this case the vector $\Omega$ turns into zero vector; hence, by definition it can not be an eigenvector.

{\sl Remark.} We have seen in Section~\ref{S-XXX} that form factor of the operator $E_m^{2,2}$ can be expressed
in terms of the $\beta$-derivative (recall that $\kappa=\eee^\beta$) of the scalar product (see \eqref{QQ}).
Knowing the zero eigenvector $\Omega$ one can calculate this derivative  explicitly. Indeed, it follows from the results of this section that if we add
to the first row of the matrix $\mathcal{N}$ all other rows multiplied by the coefficients $\Omega_j/\Omega_1$, then
the first row becomes proportional to $\kappa-1$. Then taking the $\beta$-derivative of the determinant at $\beta=0$ one should
simply differentiate this modified first row, setting $\kappa=1$ in all other rows.

\subsection{Spurious poles}

The pre-factor in \eqref{R-fin1} contains the product $t(\bmuc,\blab)$, which has poles at $\muc_j-\lab_k+c=0$ and at $\muc_j-\lab_k=0$.
On the other hand, the scalar product should not have singularities in such points. One
can check that this is really so. Let, for example, $\muc_1-\lab_1+c=0$. Then we have
\be{left-ub}
\mathcal{N}^{(\la)}(\lac_j,\lab_1)=(-1)^{a+1}r_1(\lab_1)g^{-1}(\bmuc,\lab_1)h(\blac,\lab_1)\cdot
t(\muc_1,\lac_j),
\ee
\be{right-ub}
\mathcal{N}^{(\la)}(\lac_j,\muc_1)=\kappa h(\muc_1,\blac)h(\bmuc,\muc_1)\cdot
t(\lac_j,\lab_1).
\ee
Since $t(\lac_j,\lab_1)=t(\muc_1,\lac_j)$ at $\muc_1=\lab_1-c$ (see \eqref{propert}),  we conclude that the first columns of these blocks are proportional to each other. After simple algebra
we obtain
\be{rat-u}
\left.\frac{\mathcal{N}^{(\la)}(\lac_j,\muc_1)}{\mathcal{N}^{(\la)}(\lac_j,\lab_1)}
\right|_{\muc_1=\lab_1-c}=\frac{r_3(\muc_1)}{r_1(\lab_1)},
\ee
where we have used \eqref{AEigenS-1} and  \eqref{ATEigenS-2}.

Similarly we find
\be{left-db}
\mathcal{N}^{(\muu)}(\mub_j,\lab_1)= h(\lab_1,\blab)h(\bmub,\lab_1)\cdot
t(\muc_1,\mub_j),
\ee
\be{right-db}
\mathcal{N}^{(\muu)}(\mub_j,\muc_1)=(-1)^{b+1}r_3(\muc_1)g^{-1}(\muc_1,\blab)h(\muc_1,\bmub)\cdot
t(\mub_j,\lab_1).
\ee
Using again \eqref{propert}, \eqref{AEigenS-1} and \eqref{ATEigenS-2} we obtain
\be{rat-d}
\left.\frac{\mathcal{N}^{(\muu)}(\mub_j,\muc_1)}{\mathcal{N}^{(\muu)}(\mub_j,\lab_1)}
\right|_{\muc_1=\lab_1-c}=\frac{r_3(\muc_1)}{r_1(\lab_1)},
\ee
and, hence, the first and the $(a+1)$-th columns of the matrix $\mathcal{N}$  are proportional to each other at $\muc_1-\lab_1+c=0$. Thus, the determinant vanishes.

The same effect takes place at $\muc_j=\lab_k$.

\section{Calculation of the scalar product\label{S-CSP}}

In this section we prove the representation \eqref{R-fin1}.
We start our calculations with the formula for the scalar product of  generic off-shell Bethe vectors
obtained in \cite{Res86}
 \begin{multline}\label{Resh-SP}
 \mathcal{S}_{a,b}=\sum r_1(\blab_{\so})r_1(\blac_{\st})r_3(\bmub_{\so}) r_3(\bmuc_{\st})
  f(\blac_{\so},\blac_{\st})  f(\blab_{\st},\blab_{\so})  f(\bmuc_{\st},\bmuc_{\so})
   f(\bmub_{\so},\bmub_{\st})
  \num
 \times f(\bmuc_{\so},\blac_{\so})f(\bmub_{\st},\blab_{\st}) \;Z_{a-k,n}(\blac_{\st};\blab_{\st}|\bmuc_{\so};\bmub_{\so})
 Z_{k,b-n}(\blab_{\so};\blac_{\so}|\bmub_{\st};\bmuc_{\st}).
 \end{multline}
Here the sum is taken over the partitions of the sets $\blac$, $\blab$, $\bmuc$, and $\bmub$
 \be{part-1}
 \begin{array}{ll}
 \blac=\{\blac_{\so},\;\blac_{\st}\}, &\qquad  \bmuc=\{\bmuc_{\so},\;\bmuc_{\st}\},\\
 \blab=\{\blab_{\so},\;\blab_{\st}\}, &\qquad  \bmub=\{\bmub_{\so},\;\bmub_{\st}\} .
 \end{array}
 \ee
The partitions are independent except that $\#\blab_{\so}=\#\blac_{\so}=k$ with $k=0,\dots,a$, and $\#\bmub_{\so}=\#\bmuc_{\so}=n$
with $n=0,\dots,b$.

The functions $Z_{a-k,n}$ and $Z_{k,b-n}$ are the  highest coefficients of the scalar product
\cite{Res86}. Generically $Z_{a,b}(\bar t;\bar x|\bar s; \bar y)$ depends on four sets of
variables with $\#\bar t=\#\bar x=a$ and $\#\bar s=\#\bar y=b$.
We use two equivalent representations for the highest coefficient \cite{Whe12,BelPRS12a}. The first one reads
 \be{RHC-IHC}
  Z_{a,b}(\bar t;\bar x|\bar s; \bar y)=(-1)^b\sum
 K_b(\bar s-c|\bar w_{\so})K_a(\bar w_{\st}|\bar t)
  K_b(\bar y|\bar w_{\so})f(\bar w_{\so},\bar w_{\st}).
    \ee
Here $\bar w=\{\bar s,\;\bar x\}$. The sum is taken with respect to all partitions of the set $\bar w$ into
subsets $\bar w_{\so}$ and $\bar w_{\st}$ with $\#\bar w_{\so}=b$ and $\#\bar w_{\st}=a$. The functions
$K_n$ are DWPF \eqref{K-def}.

The second representation has the following form
 \be{Al-RHC-IHC}
  Z_{a,b}(\bar t;\bar x|\bar s; \bar y)=(-1)^bf(\bar y,\bar x)f(\bar s,\bar t)
   \sum
 K_b(\bar\eta_{\st}-c|\bar y+c)K_a(\bar x|\bar\eta_{\so})K_b(\bar\eta_{\st}-c|\bar s)f(\bar\eta_{\so},\bar\eta_{\st}).
    \ee
Here $\bar\eta=\{\bar y+c,\;\bar t\}$. The sum is taken with respect to all partitions of the set $\bar\eta$ into
subsets $\bar\eta_{\so}$ and $\bar\eta_{\st}$ with $\#\bar\eta_{\so}=a$ and $\#\bar\eta_{\st}=b$.

\subsection{Summation over partitions of $\blac$ and $\bmub$}

Recall that in the framework of the generalized model we consider $r_1(\la)$ and $r_3(\muu)$ as arbitrary functions. Then every term in the
representation \eqref{Resh-SP} is labeled by a certain product $r_1(\blab_{\so})r_1(\blac_{\st})r_3(\bmub_{\so}) r_3(\bmuc_{\st})$. The terms corresponding to different partitions are labeled by different products, therefore their summation is impossible.
However, in the particular case of the scalar product of on-shell and twisted on-shell Bethe vectors one
can express the products $r_1(\blab_{\so})r_1(\blac_{\st})r_3(\bmub_{\so}) r_3(\bmuc_{\st})$ in terms of rational functions due to  \eqref{AEigenS-1}--\eqref{ATEigenS-2}. Therefore we obtain a possibility to sum up the terms corresponding
to different partitions. At the first step we calculate the sum over partitions of the sets $\blac$ and $\bmub$.

The highest coefficients $Z_{a-k,n}$ and $Z_{k,b-n}$ in \eqref{Resh-SP}  themselves are given as
sums over partitions of certain sets of their arguments. Therefore substituting explicit representations for these functions into \eqref{Resh-SP}
we create additional partitions. For example, using \eqref{RHC-IHC} we have
 \be{RHC-IHC1}
  Z_{a-k,n}(\blac_{\st};\blab_{\st}|\bmuc_{\so}; \bmub_{\so})=(-1)^n\sum
 K_n(\bmuc_{\so}-c|\bar w_{\so})K_{a-k}(\bar w_{\st}|\blac_{\st})
  K_n(\bmub_{\so}|\bar w_{\so})f(\bar w_{\so},\bar w_{\st}).
    \ee
Here $\bar w=\{\bmuc_{\so},\;\blab_{\st}\}$. The sum is taken with respect to all partitions of the set $\bar w$ into
subsets $\bar w_{\so}$ and $\bar w_{\st}$ with $\#\bar w_{\so}=n$ and $\#\bar w_{\st}=a-k$. We see that in fact
we create additional subpartitions of the subsets $\blab$ and $\bmuc$ into sub-subsets. If we use
the same formula \eqref{RHC-IHC} for the second highest coefficient $Z_{k,b-n}(\blab_{\so};\blac_{\so}|\bmub_{\st};\bmuc_{\st})$,
then we will create additional subpartitions of the subsets $\blac$ and $\bmub$. Thus, all the
subsets of variables will be divided into sub-subsets.

However if we use \eqref{Al-RHC-IHC} for $Z_{k,b-n}(\blab_{\so};\blac_{\so}|\bmub_{\st};\bmuc_{\st})$, then
 \ba{Al-RHC-IHC1}
Z_{k,b-n}(\blab_{\so};\blac_{\so}|\bmub_{\st};\bmuc_{\st})=(-1)^{b-n}f(\bmub_{\st},\blab_{\so})
f(\bmuc_{\st},\blac_{\so})  \\
  \times \sum
 K_{b-n}(\bar\eta_{\st}-c|\bmuc_{\st}+c)K_k(\lac_{\so}|\bar\eta_{\so})K_{b-n}(\bar\eta_{\st}-c|\bmub_{\st})
 f(\bar\eta_{\so},\bar\eta_{\st}).
    \end{multline}
Here $\bar\eta=\{\bmuc_{\st}+c,\;\blab_{\so}\}$, and we see that we still deal with subpartitions of the sets $\blab$ and $\bmuc$,
while the subsets of $\blac$ and $\bmub$ remain intact.

Thus, the use of different representations \eqref{RHC-IHC}, \eqref{Al-RHC-IHC} for the two highest coefficients in \eqref{Resh-SP}
allows us to keep original partitions at least for two sets of variables, namely for $\blac$ and $\bmub$.

As we have explained above, we start with the summation over partitions of the sets $\blac$ and $\bmub$.
Therefore at the first stage of the calculations it is enough to substitute \eqref{AEigenS-2} and \eqref{ATEigenS-1}
into \eqref{Resh-SP}, considering  for some time the functions $r_1(\lab)$ and $r_3(\muc)$
as free parameters.

Substituting \eqref{AEigenS-2} and \eqref{ATEigenS-1} into \eqref{Resh-SP} we obtain
 \begin{multline}\label{AResh-form-red}
 \mathcal{S}_{a,b}=\sum \kappa^{a-k}r_1(\blab_{\so}) r_3(\bmuc_{\st})
  f(\blac_{\st},\blac_{\so})  f(\blab_{\st},\blab_{\so})  f(\bmuc_{\st},\bmuc_{\so})
   f(\bmub_{\st},\bmub_{\so}) f(\bmuc_{\so},\blac_{\so})
  \num
 \times f(\bmub_{\st},\blab_{\st}) f(\bmuc,\blac_{\st})f(\bmub_{\so},\blab)\;Z_{a-k,n}(\blac_{\st};\blab_{\st}|\bmuc_{\so};\bmub_{\so})
 Z_{k,b-n}(\blab_{\so};\blac_{\so}|\bmub_{\st};\bmuc_{\st}).
 \end{multline}
Now we use \eqref{RHC-IHC1} for  $Z_{a-k,n}$, and  \eqref{Al-RHC-IHC1}
for $Z_{k,b-n}$. This gives us
 \begin{equation}\label{S-hS}
 \mathcal{S}_{a,b}= f(\bmuc,\blac)f(\bmub,\blab)\widehat{\mathcal{S}}_{a,b},
 \end{equation}
where
 \begin{multline}\label{Resh-form-red1}
 \widehat{\mathcal{S}}_{a,b}=\sum (-1)^{b}\kappa^{a-k} r_1(\blab_{\so}) r_3(\bmuc_{\st})
    f(\blab_{\st},\blab_{\so})  f(\bmuc_{\st},\bmuc_{\so})
      \num
 \times K_n(\bmuc_{\so}-c|\bar w_{\so})K_{b-n}(\bar\eta_{\st}-c|\bmuc_{\st}+c)f(\bar w_{\so},\bar w_{\st})
   f(\bar\eta_{\so},\bar\eta_{\st})
  \num
 \times \Bigl[K_n(\bmub_{\so}|\bar w_{\so})K_{b-n}(\bar\eta_{\st}-c|\bmub_{\st})f(\bmub_{\st},\bmub_{\so})\Bigr]\cdot
 \Bigl[K_{k}(\blac_{\so}|\bar\eta_{\so})K_{a-k}(\bar w_{\st}|\blac_{\st})f(\blac_{\st},\blac_{\so})\Bigr].
 \end{multline}
Recall that here $\bar w=\{\blab_{\st},\bmuc_{\so}\}$ and $\bar\eta=\{\blab_{\so},\bmuc_{\st}+c\}$. Observe that in the last line
of \eqref{Resh-form-red1} we have gathered all the terms depending of the sets $\blac$ and $\bmub$. The
sum over partitions of these sets can be calculated explicitly due to

\begin{lemma}\label{main-ident}
Let $\bar\xi$, $\bar\alpha$ and $\bar\beta$ be sets of complex variables with $\#\alpha=m_1$,
$\#\beta=m_2$, and $\#\xi=m_1+m_2$. Then
\begin{equation}\label{Sym-Part-old1}
  \sum
 K_{m_1}(\bar\xi_{\so}|\bar \alpha)K_{m_2}(\bar \beta|\bar\xi_{\st})f(\bar\xi_{\st},\bar\xi_{\so})
 = (-1)^{m_1}f(\bar\xi,\bar \alpha) K_{m_1+m_2}(\bar \alpha-c,\bar \beta|\bar\xi).
 \end{equation}
The sum is taken with respect to all partitions of the set $\bar\xi$ into
subsets $\bar\xi_{\so}$ and $\bar\xi_{\st}$ with $\#\bar\xi_{\so}=m_1$ and $\#\bar\xi_{\st}=m_2$.
Due to \eqref{Red-K} the equation \eqref{Sym-Part-old1} can be also written in the form
\begin{equation}\label{Sym-Part-old2}
  \sum
 K_{m_1}(\bar\xi_{\so}|\bar \alpha)K_{m_2}(\bar \beta|\bar\xi_{\st})f(\bar\xi_{\st},\bar\xi_{\so})
 = (-1)^{m_2}f(\bar \beta,\bar\xi) K_{m_1+m_2}(\bar\xi|\bar \alpha,\bar \beta+c).
 \end{equation}
\end{lemma}
The proof of this Lemma is given in Appendix~\ref{Pmain-ident}.

We use the equation \eqref{Sym-Part-old1} for the calculation of the sum over the partitions
of the set $\bmub$ and the equation \eqref{Sym-Part-old2} for the calculation of the sum over the
partitions of the set $\blac$. Then we obtain
 \begin{multline}\label{Part-sum1}
 \widehat{\mathcal{S}}_{a,b}=\sum (-1)^{a+b+k+n}\kappa^{a-k}r_1(\blab_{\so}) r_3(\bmuc_{\st})
    f(\blab_{\st},\blab_{\so})  f(\bmuc_{\st},\bmuc_{\so})
    f(\bar w_{\so},\bar w_{\st})
   f(\bar\eta_{\so},\bar\eta_{\st})
  \num
 \times K_n(\bmuc_{\so}-c|\bar w_{\so})K_{b-n}(\bar\eta_{\st}-c|\bmuc_{\st}+c) f(\bmub,\bar w_{\so})
 f(\bar w_{\st},\blac)\num
 \times
  K_{b}(w_{\so}-c, \bar\eta_{\st}-c|\bmub)K_{a}(\blac|\bar\eta_{\so},w_{\st}+c).
 \end{multline}

\subsection{New subsets}

To proceed further we need to introduce sub-subsets of the sets $\blab$ and $\bmuc$. We define
these sub-subsets as follows:
 \be{part-2}
 \begin{array}{ll}
 \blab_{\so}=\{\blab_{\rm i},\;\blab_{\rm iii}\},&\qquad \bar w_{\so}=\{\bmuc_{\rm i},\;\blab_{\rm ii}\} ,\\
 \blab_{\st}=\{\blab_{\rm ii},\;\blab_{\rm iv}\},&\qquad  \bar w_{\st}=\{\bmuc_{\rm iii},\;\blab_{\rm iv}\},\\
 \bmuc_{\so}=\{\bmuc_{\rm i},\;\bmuc_{\rm ii}\}, &\qquad  \bar\eta_{\so}=\{\bmuc_{\rm ii}+c,\;\blab_{\rm i}\},\\
 \bmuc_{\st}=\{\bmuc_{\rm ii},\;\bmuc_{\rm iv}\}, &\qquad  \bar\eta_{\st}=\{\bmuc_{\rm iv}+c,\;\blab_{\rm iii}\}.
 \end{array}
 \ee
The cardinalities of the introduced sub-subsets are $\#\blab_\fj=k_{\fj}$ and $\#\bmuc_\fj=n_\fj$
for $\fj={\rm i}, {\rm ii}, ...$. It is easy to
see that $k_{\rm i}+k_{\rm iii}=k$, $n_{\rm i}+n_{\rm iii}=n$, $k_{\rm ii}=n_{\rm iii}$, and $k_{\rm iii}=n_{\rm ii}$.

Due to \eqref{K-K}, \eqref{Red-K} we have
\be{sympl-KK1}
K_n(\bmuc_{\so}-c|\bar w_{\so})=K_n(\bmuc_{\rm i}-c,\bmuc_{\rm iii}-c|\bmuc_{\rm i},\blab_{\rm ii})=(-1)^nf^{-1}(\blab_{\rm ii},\bmuc_{\rm iii})
K_{n_{\rm iii}}(\blab_{\rm ii}|\bmuc_{\rm iii}).
\ee
Similarly
\begin{multline}\label{sympl-KK2}
K_{b-n}(\bar\eta_{\st}-c|\bmuc_{\st}+c)=K_{b-n}(\bmuc_{\rm iv},\blab_{\rm iii}-c|\bmuc_{\rm ii}+c,\bmuc_{\rm iv}+c)\\
=(-1)^{b-n}f^{-1}(\bmuc_{\rm ii}+c,\blab_{\rm iii})K_{n_{\rm ii}}(\bmuc_{\rm ii}+c|\blab_{\rm iii}).
\end{multline}

In terms of the introduced sub-subsets the equation \eqref{Part-sum1} takes the form
 \begin{multline}\label{SubSubsum}
 \widehat{\mathcal{S}}_{a,b}=\sum (-1)^{a+k+n}\kappa^{a-k}r_1(\blab_{\rm i})r_1(\blab_{\rm iii}) r_3(\bmuc_{\rm ii})
 r_3(\bmuc_{\rm iv})f(\bmuc_{\rm i},\blab_{\rm iv})f^{-1}(\bmuc_{\rm iv},\blab_{\rm i})\num
 \times \mathbb{F}(\blab_\fj)\mathbb{F}(\bmuc_\fj)
 f(\bmub,\bmuc_{\rm i})f(\bmub,\blab_{\rm ii})f(\bmuc_{\rm iii},\blac)f(\blab_{\rm iv},\blac)
 K_{n_{\rm iii}}(\blab_{\rm ii}|\bmuc_{\rm iii})K_{n_{\rm ii}}(\bmuc_{\rm ii}+c|\blab_{\rm iii})\num
 \times
  K_{b}(\bmuc_{\rm i}-c,\blab_{\rm ii}-c,\blab_{\rm iii}-c, \bmuc_{\rm iv}|\bmub)K_{a}(\blac|\blab_{\rm i},\bmuc_{\rm ii}+c,\bmuc_{\rm iii}+c,\blab_{\rm iv}+c),
 \end{multline}
where
\be{bbF}
\mathbb{F}(\bar z_\fj)=f(\bar z_{\rm ii},\bar z_{\rm i}) f(\bar z_{\rm ii},\bar z_{\rm iii})
f(\bar z_{\rm iv},\bar z_{\rm i}) f(\bar z_{\rm iv},\bar z_{\rm iii})
f(\bar z_{\rm ii},\bar z_{\rm iv}) f(\bar z_{\rm i},\bar z_{\rm iii}) ,
\ee
and we also have used \eqref{propert} for some functions $f$.

Now we combine sub-subsets $\blab_\fj$ and $\bmuc_\fj$ into new subsets
 \be{part-3}
 \begin{array}{ll}
 \blab_{\qo}=\{\blab_{\rm ii}\;\blab_{\rm iii}\},&\qquad \bmuc_{\qo}=\{\bmuc_{\rm ii},\;\bmuc_{\rm iii}\} ,\\
 \blab_{\qt}=\{\blab_{\rm i}\;\blab_{\rm iv}\},&\qquad  \bmuc_{\qt}=\{\bmuc_{\rm i},\;\bmuc_{\rm iv}\}.
 \end{array}
 \ee
Due to \eqref{part-2} we have $\#\blab_{\qo}=\#\bmuc_{\qo}=n_{\rm ii}+n_{\rm iii}\equiv n_{\qo}$. Observe that
these new subsets are different from the subsets used, for example, in \eqref{Resh-SP}. We use, however, the
same notation, as we deal with the sum over partitions, and therefore it does not matter how we denote separate terms
of this sum.

Then the equation \eqref{SubSubsum} can be written in the form
 \begin{equation}\label{Sub-new-part}
 \widehat{\mathcal{S}}_{a,b}=\sum
  f(\bmuc_{\qt},\blab_{\qt})f(\blab_{\qo},\blab_{\qt})f(\bmuc_{\qt},\bmuc_{\qo})
  G_{n_{\qo}}(\blab_{\qo}|\bmuc_{\qo}) \widetilde{\mathcal{L}}_a^{(\la)}(\blac|\blab_{\qt},\bmuc_{\qo})
 \widetilde{\mathcal{L}}_b^{(\muu)}(\bmub|\bmuc_{\qt},\blab_{\qo}),
 \end{equation}
where the sum is taken over the partitions of the set $\blab$ into subsets $\{\blab_{\qo},\;\blab_{\qt}\}$ and the set $\bmuc$ into subsets $\{\bmuc_{\qo},\;\bmuc_{\qt}\}$.
In this formula
 \begin{multline}\label{La}
\widetilde{\mathcal{L}}_a^{(\la)}(\blac|\blab_{\qt},\bmuc_{\qo})=
\sum_{\blab_{\qt}=\{\blab_{\rm i},\;\blab_{\rm iv}\}}K_{a}(\blac|\blab_{\rm i},\bmuc_{\qo}+c,\blab_{\rm iv}+c)
(-1)^{a-k_{\rm i}}\kappa^{a-k_{\rm i}}r_1(\blab_{\rm i})f^{-1}(\bmuc,\blab_{\rm i})\\
 \times f(\blab_{\rm iv},\blab_{\rm i}) f(\bmuc_{\qo},\blab_{\rm i})f(\blab_{\rm iv},\blac)f(\bmuc_{\qo},\blac),
  \end{multline}
and
 \begin{multline}\label{Lb}
 \widetilde{\mathcal{L}}_b^{(\muu)}(\bmub|\bmuc_{\qt},\blab_{\qo})=
 \sum_{\bmuc_{\qt}=\{\bmuc_{\rm i},\;\bmuc_{\rm iv}\}}K_{b}(\bmuc_{\rm i}-c,\blab_{\qo}-c, \bmuc_{\rm iv}|\bmub)
 (-1)^{b-n_{\rm iv}}r_3(\bmuc_{\rm iv}) f^{-1}(\bmuc_{\rm iv},\blab)
 \num
 \times f(\bmuc_{\rm iv},\bmuc_{\rm i})f(\bmuc_{\rm iv},\blab_{\qo})f(\bmub,\bmuc_{\rm i})f(\bmub,\blab_{\qo}).
  \end{multline}
Finally, the function $G_{n_{\qo}}(\blab_{\qo}|\bmuc_{\qo})$ is given by
 \be{Gn1}
 G_{n_{\qo}}(\blab_{\qo}|\bmuc_{\qo})=
 \sum_{\blab_{\qo}=\{\blab_{\rm ii},\;\blab_{\rm iii}\}\atop{\bmuc_{\qo}=\{\bmuc_{\rm ii},\;\bmuc_{\rm iii}\}}}
 \kappa^{-k_{\rm iii}}\hat r_1(\blab_{\rm iii})\hat r_3(\bmuc_{\rm ii})
 f(\bmuc_{\rm iii},\bmuc_{\rm ii})f(\blab_{\rm iii},\blab_{\rm ii})
  K_{n_{\rm iii}}(\blab_{\rm ii}|\bmuc_{\rm iii})K_{n_{\rm ii}}(\bmuc_{\rm ii}+c|\blab_{\rm iii}),
  \ee
 where
\be{hr1}
\hat r_1(\lab_j)=r_1(\lab_j)\prod_{k=1\atop{k\ne j}}^a\frac{f(\lab_k,\lab_j)}{f(\lab_j,\lab_k)}\prod_{m=1}^bf^{-1}(\mub_m,\lab_j),
\ee
and
\be{hr3}
\hat r_3(\muc_j)=r_3(\muc_j)\prod_{k=1\atop{k\ne j}}^b\frac{f(\muc_j,\muc_k)}{f(\muc_k,\muc_j)}\prod_{\ell=1}^af^{-1}(\muc_j,\lac_\ell).
\ee

We stress that the representation \eqref{Sub-new-part} follows from \eqref{SubSubsum}
without any additional transforms. One can check that the straightforward substitution
of \eqref{La}--\eqref{hr3} into \eqref{Sub-new-part} gives exactly \eqref{SubSubsum}. Thus, at the
second step of our calculations we have not sum up any sum over partitions in \eqref{SubSubsum};
however, we have factorized the original sum into several subsums.

\subsection{Summation over subpartitions of $\blab$ and $\bmuc$}

The functions $\widetilde{\mathcal{L}}_a^{(\la)}$, $\widetilde{\mathcal{L}}_b^{(\muu)}$, $G_{n_{\qo}}$, which appear in
\eqref{Sub-new-part} are given in terms of sums over partitions. These sums can be computed
explicitly. We begin with the functions $\widetilde{\mathcal{L}}_a^{(\la)}$ and $\widetilde{\mathcal{L}}_b^{(\muu)}$.

\begin{lemma}\label{Long-Det}
Let $\bar w$ and $\bar\xi$ be two sets of generic complex numbers with $\#\bar w=\#\bar\xi=m$. Let
also $C_1(w)$ and $C_2(w)$ be two arbitrary functions of a complex variable $w$. Then
\begin{multline}\label{SumDet1}
\sum K_m(\bar w_{\qo}-c, \bar w_{\qt}|\bar \xi)f(\bar \xi, \bar w_{\qo})f(\bar w_{\qt},\bar w_{\qo})
C_1(\bar w_{\so})C_2(\bar w_{\st})\num
=\Delta'_m(\bar\xi)\Delta_m(\bar w)
\det_m\Bigl(C_2(w_k)t(w_k,\xi_j)h(w_k,\bar\xi)+(-1)^m C_1(w_k)t(\xi_j,w_k)h(\bar\xi,w_k)\Bigr),
\end{multline}
and
\begin{multline}\label{SumDet2}
\sum K_m(\bar \xi|\bar w_{\qo}, \bar w_{\qt}+c)f(\bar w_{\qt},\bar \xi)f(\bar w_{\qt},\bar w_{\qo})
C_1(\bar w_{\so})C_2(\bar w_{\st})\num
=\Delta'_m(\bar\xi)\Delta_m(\bar w)
\det_m\Bigl(C_1(w_k)t(\xi_j,w_k)h(\bar\xi,w_k) +(-1)^m C_2(w_k) t(w_k,\xi_j)h(w_k,\bar\xi)\Bigr).
\end{multline}
Here the sums are taken over all possible partitions of the set $\bar w$ into subsets $\bar w_{\qo}$
and $\bar w_{\qt}$. Recall that $\Delta'_m$ and $\Delta_m$ are  given by \eqref{def-Del}.

\end{lemma}
The proof is given in Appendix~\ref{PLong-Det}.

We see that we can apply Lemma~\ref{Long-Det} in order to obtain determinant representations
for $\widetilde{\mathcal{L}}_a^{(\la)}$ \eqref{La} and $\widetilde{\mathcal{L}}_b^{(\mu)}$ \eqref{Lb}. For instance,
if we set in \eqref{SumDet2}: $m=a$, $\bar \xi=\blac$, $\bar w=\{\blab_{\qt},\;\bmuc_{\qo}\}$ and
 \be{C1C2}
 C_1(w)=r_1(w)f^{-1}(\bmuc,w), \qquad C_2(w)=-\kappa,
 \ee
then we obtain the equation \eqref{La}. Indeed, in this case one has $C_1(\muc_j)=0$ due to the product
$f^{-1}(\bmuc,w)$ in \eqref{C1C2}. Hence, we automatically have $\bmuc_{\qo}\subset \bar w_{\qt}$, otherwise
the corresponding contribution to the sum vanishes. This means that
when splitting the set $\bar w=\{\blab_{\qt},\;\bmuc_{\qo}\}$
into two subsets we actually should consider only the partitions of the set $\blab_{\qt}$ into $\blab_{\rm i}$ and
$\blab_{\rm iv}$, as we have in \eqref{La}.   We
obtain
 \be{ALa}
 \widetilde{\mathcal{L}}_a^{(\la)}(\blac|\bar w) =\Delta'_a(\blac)\Delta_a(\bar w)
 \det_a\widetilde{\mathcal{N}}^{(\la)}(\lac_j,w_k),
 \ee
with
 \be{ANa}
 \widetilde{\mathcal{N}}^{(\la)}(\lac_j,w_k)=r_1(w_k)t(\lac_j,w_k)h(\blac,w_k)f^{-1}(\bmuc,w_k) -\kappa(-1)^a
  t(w_k,\lac_j)h(w_k,\blac),
 \ee
and $\bar w=\{\blab_{\qt},\;\bmuc_{\qo}\}$.
Similarly the sum in \eqref{Lb} can be calculated via \eqref{SumDet1}. We have
 \be{ALb}
 \widetilde{\mathcal{L}}_b^{(\muu)}(\bmub|\bar w) =\Delta'_b(\bmub)\Delta_b(\bar w)
 \det_b\widetilde{\mathcal{N}}^{(\muu)}(\mub_j,w_k),
 \ee
with
 \be{ANb}
 \widetilde{\mathcal{N}}^{(\muu)}(\mub_j,w_k)=r_3(w_k) t(w_k,\mub_j)h(w_k,\bmub)f^{-1}(w_k,\blab) -(-1)^b
 t(\mub_j,w_k)h(\bmub,w_k ),
 \ee
and $\bar w=\{\blab_{\qo},\;\bmuc_{\qt}\}$.

Observe that up to now we did not use the constraints \eqref{AEigenS-1}, \eqref{ATEigenS-2}. However, we should use
them for the calculation of $G_{n_{\qo}}(\blab_{\qo}|\bmuc_{\qo})$ \eqref{Gn1}. Then $\hat r_1(\lab_j)=1$,
$\hat r_3(\muc_j)=\kappa$, and \eqref{Gn1} turns into
 \be{Gn1-red}
 G_{n_{\qo}}(\blab_{\qo}|\bmuc_{\qo})=
 \sum_{\blab_{\qo}=\{\blab_{\rm ii},\;\blab_{\rm iii}\}\atop{\bmuc_{\qo}=\{\bmuc_{\rm ii},\;\bmuc_{\rm iii}\}}}
 f(\bmuc_{\rm iii},\bmuc_{\rm ii})f(\blab_{\rm iii},\blab_{\rm ii})   K_{n_{\rm iii}}(\blab_{\rm ii}|\bmuc_{\rm iii})
 K_{n_{\rm ii}}(\bmuc_{\rm ii}+c|\blab_{\rm iii}).
  \ee
For this calculation, we need one more lemma:

\begin{lemma}\label{Wau}
Let $\bar\alpha$ and $\bar\beta$ be two sets of generic complex numbers with $\#\bar\alpha=\#\bar\beta=m$.
Then
\begin{equation}\label{Ident-G}
\sum_{\bar\alpha=\{\bar\alpha_{\qo},\;\bar\alpha_{\qt}\}\atop{\bar\beta=\{\bar\beta_{\qo},\;\bar\beta_{\qt}\}}}
 f(\bar\beta_{\qt},\bar\beta_{\qo})f(\bar\alpha_{\qo},\bar\alpha_{\qt})   K_{m_{\qo}}(\bar\beta_{\qo}|\bar\alpha_{\qo})K_{m_{\qt}}(\bar\alpha_{\qt}+c|\bar\beta_{\qt})
  =(-1)^m t(\bar\alpha,\bar\beta)h(\bar\alpha,\bar\alpha)h(\bar\beta,\bar\beta),
       \end{equation}
where the sum is taken over all possible partitions of the sets $\bar\alpha$ and $\bar\beta$
with $\#\bar\alpha_{\qo}=\#\bar\beta_{\qo}=m_{\qo}$ and $\#\bar\alpha_{\qt}=\#\bar\beta_{\qt}=m_{\qt}$.
\end{lemma}
The proof is given in Appendix~\ref{Wonderful}. Obviously,  equation \eqref{Gn1-red} coincides with \eqref{Ident-G}
after appropriate identification of the subsets. Thus,
 \be{Gn1-red-answ}
 G_{n_{\qo}}(\blab_{\qo}|\bmuc_{\qo})=(-1)^{n_{\qo}}t(\bmuc_{\qo},\blab_{\qo})h(\blab_{\qo},\blab_{\qo})
 h(\bmuc_{\qo},\bmuc_{\qo}).
 \ee
Substituting this into \eqref{Sub-new-part} we arrive at
 \begin{multline}\label{Sub-new-part-red}
 \widehat{\mathcal{S}}_{a,b}=\sum_{\blab=\{\blab_{\qo},\;\blab_{\qt}\}\atop{\bmuc=\{\bmuc_{\qo},\;\bmuc_{\qt}\}}}
  (-1)^{n_{\qo}}f(\bmuc_{\qt},\blab_{\qt})f(\blab_{\qo},\blab_{\qt})f(\bmuc_{\qt},\bmuc_{\qo})
 t(\bmuc_{\qo},\blab_{\qo})h(\blab_{\qo},\blab_{\qo})
 h(\bmuc_{\qo},\bmuc_{\qo})\num
 \times \widetilde{\mathcal{L}}_a^{(\la)}(\blac|\blab_{\qt},\bmuc_{\qo})
 \widetilde{\mathcal{L}}_b^{(\muu)}(\bmub|\bmuc_{\qt},\blab_{\qo}),
 \end{multline}
where $\widetilde{\mathcal{L}}_a^{(\la)}$ and $\widetilde{\mathcal{L}}_b^{(\muu)}$ are given by \eqref{ALa} and \eqref{ALb} respectively.

Note that although we have used already  \eqref{AEigenS-1} and \eqref{ATEigenS-2}, we still can keep the explicit
dependence on $r_1(\lab_k)$ and $r_3(\muc_k)$ in the formulas \eqref{ANa} and \eqref{ANb} for the matrices $\widetilde{\mathcal{N}}^{(\la)}$ and $\widetilde{\mathcal{N}}^{(\muu)}$. Of course, we  cannot consider $r_1(\lab_k)$ and $r_3(\muc_k)$ as arbitrary functional
parameters anymore, but on the other hand it is not necessary to replace these functions immediately via the constraints
\eqref{AEigenS-1} and \eqref{ATEigenS-2}.  On the contrary, it is useful to keep \eqref{ANa} and \eqref{ANb} in
their present form, since it is more convenient for taking the limit $\blac\to\blab$ and $\bmuc\to\bmub$, as we have
seen in Section~\ref{S-Norm}.

\subsection{Final summation over partitions of $\blab$ and $\bmuc$}

It remains to sum up the equation \eqref{Sub-new-part-red} into a single determinant. For this we introduce
new matrices
 \be{AN-def1}
 \begin{array}{l}
  {\dis\mathcal{N}^{(\la)}(\lac_j,w_k) =(-1)^{a-1}\widetilde{\mathcal{N}}^{(\la)}(\lac_j,w_k) h(\bmuc,w_k) ,}\num
 {\dis\mathcal{N}^{(\muu)}(\mub_j,w_k) =(-1)^{b-1}\widetilde{\mathcal{N}}^{(\muu)}(\mub_j,w_k)h(w_k,\blab).}
 \end{array}
 \ee
Respectively we define
 \be{AL-def1}
 \begin{array}{l}
 {\dis\mathcal{L}_a^{(\la)}(\blac|\bar w) =\Delta'_a(\blac)\Delta_a(\bar w)
 \det_a\mathcal{N}^{(\la)}(\lac_j,w_k)=h(\bmuc,\bar w)\widetilde{\mathcal{L}}_a^{(\la)}(\blac|\bar w),}\num
  {\dis\mathcal{L}_b^{(\muu)}(\bmub|\bar x) =\Delta'_b(\bmub)\Delta_b(\bar w)
 \det_b\mathcal{N}^{(\muu)}(\mub_j,w_k)=h(\bar w,\blab)\widetilde{\mathcal{L}}_b^{(\muu)}(\bmub|\bar w).}
 \end{array}
 \ee

Observe that if we set $w_k=\lab_k$ in \eqref{AN-def1} and use \eqref{AEigenS-1}, then  $\mathcal{N}^{(\la)}$ and
$\mathcal{N}^{(\muu)}$ turn into \eqref{R-P11} and \eqref{R-P21} respectively. Setting in \eqref{AN-def1} $w_k=\muc_k$  and using \eqref{ATEigenS-2} we obtain the blocks \eqref{R-P12} and \eqref{R-P22}.

It is easy  to see that in terms of the introduced objects  equation  \eqref{Sub-new-part-red} takes the form
 \begin{equation}\label{Pre-fin}
 \widehat{\mathcal{S}}_{a,b}= t(\bmuc,\blab)\sum_{\blab=\{\blab_{\qo},\;\blab_{\qt}\}\atop{\bmuc=\{\bmuc_{\qo},\;\bmuc_{\qt}\}}}
  (-1)^{n_{\qo}}\frac{g(\bmuc_{\qt},\bmuc_{\qo})g(\blab_{\qo},\blab_{\qt})}
 {g(\bmuc_{\qo},\blab_{\qt})g(\bmuc_{\qt},\blab_{\qo})}
 \mathcal{L}_a^{(\la)}(\blac|\blab_{\qt},\bmuc_{\qo})
 \mathcal{L}_b^{(\muu)}(\bmub|\bmuc_{\qt},\blab_{\qo}).
 \end{equation}
The last sum, in fact, is an expansion of the determinant of an $(a+b)\times(a+b)$ matrix.

\begin{lemma}\label{BM}
Let $\mathcal{N}$ be an $(a+b)\times(a+b)$ matrix with block structure
 \begin{equation}\label{block-matrix}
\mathcal{N}=\left(\begin{array}{cc}
 {\dis \mathcal{N}^{(\la)}(\lac_j,\lab_k)}&
 {\dis \mathcal{N}^{(\la)}(\lac_j,\muc_k)}\num
 {\dis \mathcal{N}^{(\muu)}(\mub_j,\lab_k)}&
 {\dis \mathcal{N}^{(\muu)}(\mub_j,\muc_k)}
 \end{array}\right).
 \end{equation}
Then
 \begin{equation}\label{fin0}
 \widehat{\mathcal{S}}_{a,b}= t(\bmuc,\blab)
 \Delta'_a(\blac) \Delta'_b(\bmub)
 {\Delta}_a(\blab){\Delta}_b(\bmuc)\det_{a+b}\mathcal{N}.
 \end{equation}
\end{lemma}
{\sl Proof.} Let $\bar w=\{\blab,\bmuc\}$, that is,
 \be{w-lm}
 w_1,\dots,w_{a+b}=\lab_1,\dots,\lab_{a},\muc_1,\dots,\muc_{b}.
 \ee
Then the matrix $\mathcal{N}$ can be written as a matrix
consisting of two block-rows:
\be{mat-mat}
\mathcal{N}=
 \left(\begin{array}{c}
 {\dis \mathcal{N}^{(\la)}(\lac_j,w_k)}\num
 {\dis \mathcal{N}^{(\muu)}(\mub_j,w_k)}
 \end{array}\right).
\ee
Let us develop $\det_{a+b}\mathcal{N}$ with respect to
the first block-row:
 \begin{multline}\label{lemma-prove1}
\Delta'_a(\blac) \Delta'_b(\bmub) {\Delta}_a(\blab){\Delta}_b(\bmuc)
 \det_{a+b}\mathcal{N}=
 \Delta'_a(\blac) \Delta'_b(\bmub) {\Delta}_a(\blab){\Delta}_b(\bmuc)\num
 \times
 \sum_{\bar w=\{\bar w^{\so},\;\bar w^{\st}\}} (-1)^{P_{\so,\st}} \det_a \mathcal{N}^{(\la)}(\lac_j,w^{\so}_k)
 \det_b\mathcal{N}^{(\muu)}(\mub_j,w^{\st}_k).
\end{multline}
The sum in \eqref{lemma-prove1} is taken over all partitions of $\bar w$
with $\#\bar w^{\so}=a$. We also have denoted by $w^{\so}_k$  (respectively by $w^{\st}_k$) the $k$th element of the subset  $w^{\so}$
 (respectively $w^{\st}$).  The symbol ${P_{\so,\st}}$ means the parity of the permutation that maps the sequence $\{\bar w^{\so},\bar w^{\st}\}$ into the ordered set $\bar w$.  In particular,
 \be{Del-Del}
 (-1)^{P_{\so,\st}}= \frac{\Delta_a(\bar w^{\so})\Delta_b(\bar w^{\st}) g(\bar w^{\st},\bar w^{\so})}
 {\Delta_{a+b}(\bar w)}.
 \ee
Substituting \eqref{Del-Del}  into \eqref{lemma-prove1} and using  $\Delta_{a+b}(\bar w)=\Delta_a(\blab)
\Delta_b(\bmuc)g(\bmuc,\blab)$
we obtain
 \begin{equation}\label{lemma-prove2}
\Delta'_a(\blac) \Delta'_b(\bmub) {\Delta}_a(\blab){\Delta}_b(\bmuc)
 \det_{a+b}\mathcal{N}
  = \sum_{\bar w=\{\bar w^{\so},\;\bar w^{\st}\}} \frac{g(\bar w^{\st},\bar w^{\so})}{g(\bmuc,\blab)}
 \mathcal{L}_{a}^{(\la)}(\blac|\bar w^{\so})
 \mathcal{L}_{b}^{(\muu)}(\bmub|\bar w^{\st}) ,
\end{equation}
where we have used the definitions \eqref{AL-def1}.  Now we set
$\bar w^{\so}=\{\blab_{\st},\bmuc_{\so}\}$ and $\bar w^{\st}= \{\blab_{\so},\bmuc_{\st}\}$, and we arrive at
 \begin{multline}\label{lemma-prove3}
\Delta'_a(\blac) \Delta'_b(\bmub) {\Delta}_a(\blab){\Delta}_b(\bmuc)
 \det_{a+b}\mathcal{N}= \sum_{\blab=\{\blab_{\qo},\;\blab_{\qt}\}\atop{\bmuc=\{\bmuc_{\qo},\;\bmuc_{\qt}\}}}
  (-1)^{n_{\qo}}\frac{g(\bmuc_{\qt},\bmuc_{\qo})g(\blab_{\qo},\blab_{\qt})}
 {g(\bmuc_{\qo},\blab_{\qt})g(\bmuc_{\qt},\blab_{\qo})}\num
 \times\mathcal{L}_a^{(\la)}(\blac|\blab_{\qt},\bmuc_{\qo})
 \mathcal{L}_b^{(\muu)}(\bmub|\bmuc_{\qt},\blab_{\qo}),
 \end{multline}
which ends the proof.\hfill $\square$

Finally, using \eqref{S-hS} we arrive at the representation \eqref{R-fin1}.

\section*{Conclusion}

We have conjectured in our recent paper \cite{BelPRS12a} that in the $SU(3)$ case a single determinant representation for
the highest coefficient $Z_{a,b}$ (see \eqref{RHC-IHC}, \eqref{Al-RHC-IHC})  does not exist. It follows from
this conjecture that there is not a single determinant representation for the scalar product of on-shell and generic off-shell
Bethe vectors. We have seen, however, that it is possible to derive the determinant representation for the scalar product, if 
we deal with a twisted on-shell Bethe vector that can be considered as a
particular case of an off-shell Bethe vector. Moreover, this particular case is important from the viewpoint of its
application to the calculation of form factors of local operators. Therefore a natural development of our method would be
to study special cases of scalar products involving specific off-shell vectors. In particular, one can consider
off-shell vectors arising as the result of the action of the monodromy matrix entries $T_{jk}(w)$ on on-shell vectors.
The corresponding scalar products then can be used for the calculation of form factors and correlation functions.

It is worth mentioning that following this route one has to solve an additional problem.
Namely, it is necessary to express the action of the matrix elements $T_{jk}(w)$ on Bethe vectors in terms of a linear combination of off-shell vectors. This problem has a well known solution for $\frak{gl}_2$-based models (see e.g. \cite{BogIK93L,FadLH96}).
However,  in the case of higher rank algebras this question has not been studied yet.
We will give a solution of this problem in our forthcoming publication \cite{BelPRS12c}.

\section*{Acknowledgements}
The work of SZP was supported in part by RFBR grant 11-01-00980-a, grant
of Scientific Foundation of NRU HSE ╣ 12-09-0064 and  grant of
FASI RF 14.740.11.0347. ER was supported by ANR Project 
DIADEMS (Programme Blanc ANR SIMI1 2010-BLAN-0120-02).
NAS was  supported by the Program of RAS Basic Problems of the Nonlinear Dynamics,
RFBR-11-01-00440, RFBR-11-01-12037-ofi-m, SS-4612.2012.1.

\appendix

\section{The properties of the DWPF\label{A-IHC}}

The DWPF is  symmetric function of $x_1,\dots,x_n$ and symmetric function of $y_1,\dots,y_n$.
It behaves as $1/x_n$ (respectively $1/y_n$) as $x_n\to\infty$ (respectively $y_n\to\infty$) for other variables fixed.
It has simple poles at $x_j=y_k$. The behavior of $K_n$ near  these poles can be expressed in terms of $K_{n-1}$,
 \be{Rec-Ky}
\Bigl. K_n(\bar x|\bar y)\Bigr|_{x_n\to y_n}= g(x_n,y_n)
f(y_n,\bar y')f(\bar x',x_n)\cdot K_{n-1}(\bar x'|\bar y')+ reg,
\ee
where $\bar x'=\bar x\setminus x_n$ and $\bar y'=\bar y\setminus y_n$ and $reg$ means the regular part at $x_n\to y_n$.

One can also easily check that the DWPF possesses the properties
 \be{K-K}
K_{n+1}(\bar x, z-c|\bar y, z)=K_{n+1}(\bar x, z|\bar y, z+c)= - K_{n}(\bar x|\bar y),
\ee
 \be{Red-K}
K_{n}(\bar x-c|\bar y)=K_{n}(\bar x|\bar y+c)= (-1)^n f^{-1}(\bar y,\bar x) K_{n}(\bar y|\bar x).
\ee

\section{Proof of the Lemma~\eqref{main-ident}\label{Pmain-ident}}

The proof will be given by the induction over $m_1$. For $m_1=0$ the identity
\eqref{Sym-Part-old1} is obviously valid. Suppose that it is valid for some $m_1-1$ and
consider the l.h.s. of \eqref{Sym-Part-old1} as a function of $\alpha_1$,
\begin{equation}\label{Fal}
 F_l(\alpha_1)= \sum
 K_{m_1}(\bar\xi_{\so}|\bar \alpha)K_{m_2}(\bar \beta|\bar\xi_{\st})f(\bar\xi_{\st},\bar\xi_{\so}).
 \end{equation}
This function
decreases as $\alpha_1\to\infty$ and it has poles at $\alpha_1=\xi_k$, $k=1,\dots,m_1+m_2$.
Consider the behavior of $F_l(\alpha_1)$ near the pole at $\alpha_1=\xi_1$ (due to the symmetry of $F_l(\alpha_1)$ over $\bar\xi$
its properties near other poles are similar).

Obviously the pole at $\alpha_1=\xi_1$ occurs if and only if $\xi_1\in\bar\xi_{\so}$. Let $\bar\xi'_{\so}=
\bar\xi_{\so}\setminus\xi_1$, \ $\bar\xi'=\bar\xi\setminus\xi_1$ and $\bar\alpha'=\bar\alpha\setminus\alpha_1$.
Then using \eqref{Rec-Ky} we have
\begin{equation}\label{Fa-res}
 \Bigl.F_l(\alpha_1)\Bigr|_{\alpha_1\to\xi_1}= {\sum}' g(\xi_1,\alpha_1)f(\alpha_1,\bar\alpha')
 f(\bar\xi'_{\so},\xi_1)
 K_{m_1-1}(\bar\xi'_{\so}|\bar \alpha')K_{m_2}(\bar \beta|\bar\xi_{\st})
 f(\bar\xi_{\st},\bar\xi'_{\so})f(\bar\xi_{\st},\xi_1),
 \end{equation}
where the symbol $\sum'$ means that the summation is taken over the partitions of the set $\bar\xi'$ into
subsets $\bar\xi'_{\so}$ and $\bar\xi_{\st}$. The factors $g(\xi_1,\alpha_1)f(\alpha_1,\bar\alpha')$ and
$f(\bar\xi'_{\so},\xi_1)f(\bar\xi_{\st},\xi_1)=f(\bar\xi',\xi_1)$ can be moved out of the sum over these partitions.
Applying the induction assumption for the remaining sum we arrive at
\begin{equation}\label{Fal-res1}
 \Bigl.F_l(\alpha_1)\Bigr|_{\alpha_1\to\xi_1}= (-1)^{m_1-1} g(\xi_1,\alpha_1)f(\alpha_1,\bar\alpha')f(\bar\xi',\xi_1)
 f(\bar\xi',\bar\alpha') K_{m_1+m_2-1}(\bar \alpha'-c,\bar\beta|\bar\xi').
 \end{equation}

Consider now the rhs of \eqref{Sym-Part-old1} as a function of $\alpha_1$,
\begin{equation}\label{Fa}
 F_r(\alpha_1)=(-1)^{m_1}f(\bar\xi,\bar\alpha) K_{m_1+m_2}(\bar \alpha-c,\bar\beta|\bar\xi).
 \end{equation}
Here the pole at $\alpha_1=\xi_1$ occurs only in the prefactor $f(\bar\xi,\bar\alpha)$. Using \eqref{K-K} we
immediately obtain
\begin{equation}\label{Far-res1}
 \Bigl.F_r(\alpha_1)\Bigr|_{\alpha_1\to\xi_1}= (-1)^{m_1-1} g(\xi_1,\alpha_1)f(\alpha_1,\bar\alpha')f(\bar\xi',\xi_1)
 f(\bar\xi',\bar\alpha') K_{m_1+m_2-1}(\bar \alpha'-c,\bar\beta|\bar\xi').
 \end{equation}
Thus, the residue of the difference $F_l(\alpha_1)-F_r(\alpha_1)$ at $\alpha_1=\xi_1$ vanishes. Due to the
symmetry over $\bar\xi$ the difference $F_l(\alpha_1)-F_r(\alpha_1)$ has no poles at $\alpha_1=\xi_k$,
$k=1,\dots,m_1+m_2$. Hence, this is a holomorphic function of $\alpha_1$ in the whole complex plane. Since
this function decreases at infinity, we conclude that $F_l(\alpha_1)\equiv F_r(\alpha_1)$.

The identity \eqref{Sym-Part-old2} follows from \eqref{Sym-Part-old1} due to \eqref{Red-K}. \hfill $\square$

\section{Proof of the Lemma~\eqref{Long-Det}\label{PLong-Det}}

We shall prove only the identity \eqref{SumDet1}, as the proof of  \eqref{SumDet2} is identical.
Let us denote by $M$ the matrix in the rhs of \eqref{SumDet1}. Obviously $\det_m M$ is a linear
function of every $C_1(w_k)$ and every $C_2(w_k)$, where $k=1,\dots,m$. This function can
be presented in the form
 \be{M-pres}
 \det_m M=\sum C_1(\bar w^{\qo})C_2(\bar w^{\qt})A(\bar w^{\qo},\bar w^{\qt}),
 \ee
where $A(\bar w^{\qo},\bar w^{\qt})$ does not depend on $C_1$ and $C_2$. The sum is taken over all
possible partitions of the set $\bar w$ into subsets $\bar w^{\qo}$ and $\bar w^{\qt}$ with
$\#\bar w^{\qo}=n$, $\#\bar w^{\qt}=m-n$, $n=0,\dots,m$. In order to find the coefficient $A(\bar w^{\qo},\bar w^{\qt})$
one should simply set $C_1(\bar w^{\qt})=0$ and $C_2(\bar w^{\qo})=0$ in the rhs of \eqref{SumDet1}.
We obtain
 \be{A-formul}
 A(\bar w^{\qo},\bar w^{\qt})=(-1)^{P_{\qo,\qt}}(-1)^{nm} h(\bar\xi,\bar w^{\qo})h(w^{\qt},\bar\xi)
  \det_m\bigl(t(\xi_j,w^{\qo}_k)
 \;\mid\; t(w^{\qt}_k,\xi_j)\bigr).
 \ee
Here $P_{\qo,\qt}$ is the parity of the permutation mapping the set $\{\bar w^{\qo},\; \bar w^{\qt}\}$ into
the ordered set $\bar w$ (recall that in every subset the elements are ordered in the natural order). The
determinant in \eqref{A-formul} consists of two parts. The first $n$ columns are associated with the
parameters $\bar w^{\qo}$, while the last $m-n$ columns are associated with the
parameters $\bar w^{\qt}$. Therefore, just like in \eqref{lemma-prove1} we have  denoted by $w^{\so}_k$  (respectively by $w^{\st}_k$) the $k$-th element of the subset  $w^{\so}$ (respectively $w^{\st}$).

Thus, the rhs of \eqref{SumDet1} has the following representation
\begin{multline}\label{RHS-1}
\Delta'_m(\bar\xi)\Delta_m(\bar w)\det_m M=
\Delta'_m(\bar\xi)\Delta_m(\bar w)
\sum (-1)^{P_{\qo,\qt}}(-1)^{nm} C_1(\bar w^{\qo})C_2(\bar w^{\qt})\\
\times h(\bar\xi,\bar w^{\qo})h(\bar w^{\qt},\bar\xi)
  \det_m\bigl(t(\xi_j,w^{\qo}_k)
 \;\mid\; t(w^{\qt}_k,\xi_j)\bigr).
\end{multline}

On the other hand in the lhs of \eqref{SumDet1} it is enough to write  the DWPF
$K_m(\bar w^{\qo}-c, \bar w^{\qt}|\bar \xi)$ explicitly. We have
\begin{multline}\label{K-expl}
K_m(\bar w^{\qo}-c, \bar w^{\qt}|\bar \xi)
=\Delta'_m(\bar\xi)\Delta_n(\bar w^{\qo})\Delta_{m-n}(\bar w^{\qt})
h^{-1}(\bar w^{\qt},\bar w^{\qo})\num
\times g^{-1}(\bar w^{\qo},\bar\xi)h(\bar w^{\qt},\bar\xi)
  \det_m\bigl(t(\xi_j,w^{\qo}_k)
 \;\mid\; t(w^{\qt}_k,\xi_j)\bigr).
\end{multline}
Here we have used \eqref{propert}. Hence,
\begin{multline}\label{K-expl1}
K_m(\bar w^{\qo}-c, \bar w^{\qt}|\bar \xi)f(\bar \xi, \bar w^{\qo})f(\bar w^{\qt},\bar w^{\qo})
=\Delta'_m(\bar\xi)\Delta_n(\bar w^{\qo})\Delta_{m-n}(\bar w^{\qt})
g(\bar w^{\qt},\bar w^{\qo})\num
\times(-1)^{nm} h(\bar\xi,\bar w^{\qo})h(\bar w^{\qt},\bar\xi)
  \det_m\bigl(t(\xi_j,w^{\qo}_k)
 \;\mid\; t(w^{\qt}_k,\xi_j)\bigr).
\end{multline}
Finally, using
\be{2del}
\Delta_n(\bar w^{\qo})\Delta_{m-n}(\bar w^{\qt})
g(\bar w^{\qt},\bar w^{\qo})=(-1)^{P_{\qo,\qt}}\Delta(\bar w),
\ee
and substituting this into the rhs of \eqref{K-expl1} we arrive at \eqref{RHS-1}.\hfill $\square$

\section{Proof of Lemma~\ref{Wau} \label{Wonderful}}

Let us denote by $\Lambda^{(l)}_m(\bar\alpha|\bar\beta)$ and $\Lambda^{(r)}_m(\bar\alpha|\bar\beta)$
the lhs and the rhs of \eqref{Ident-G} respectively. Then these functions possess the
following properties:
\begin{itemize}
 \item they are rational functions of $\bar\alpha$ and $\bar\beta$;

\item they are symmetric functions of $\bar\alpha$ and symmetric functions of $\bar\beta$;

\item they vanish if some $\alpha_j$ or $\beta_j$ goes to infinity;

\item they have poles only at $\alpha_j=\beta_k$ and $\alpha_j+c=\beta_k$;

\item
 \be{G1}
 \Lambda^{(l)}_1(\alpha|\beta)=\Lambda^{(r)}_1(\alpha|\beta)=-t(\alpha,\beta).
 \ee
\end{itemize}

Therefore it is enough to check that the residues of both
functions at $\alpha_1=\beta_1$ and $\alpha_1+c=\beta_1$ coincide. In fact, we can
prove that both functions possess the same recurrence in the mentioned poles. Then,
similarly to the proof of Lemma~\ref{main-ident} we can use the induction over $m$.

It is straightforward to establish the following recursions:
\begin{align}\label{recursL1}
 \Bigl. \Lambda^{(r)}_m(\bar\alpha|\bar\beta)\Bigr|_{\alpha_1\to \beta_1} &=
 -g(\alpha_1, \beta_1)f(\bar\beta',\beta_1)f(\alpha_1,\bar\alpha')\cdot \Lambda^{(r)}_{m-1}(\bar\alpha'|\bar\beta'),\\
 \Bigl. \Lambda^{(r)}_m(\bar\alpha|\bar\beta)\Bigr|_{\alpha_1+c \to \beta_1} &=
 h^{-1}(\alpha_1, \beta_1)f(\beta_1,\bar\beta')f(\bar\alpha',\alpha_1)\cdot \Lambda^{(r)}_{m-1}(\bar\alpha'|\bar\beta'),
 \label{recursL2}
\end{align}
where $\bar\alpha'=\bar\alpha\setminus\alpha_1$
and $\bar\beta'=\bar\beta\setminus\beta_1$. It is not difficult to see that $\Lambda^{(l)}_m(\bar\alpha|\bar\beta)$
has the same recursion properties.

Consider, for example, the pole at $\alpha_1=\beta_1$. This pole appears if and only if
$\alpha_1\in\bar\alpha_{\qo}$ and $\beta_1\in\bar\beta_{\qo}$. Let $\bar\alpha_{{\qo}'}=\bar\alpha_{\qo}\setminus\alpha_1$
and $\bar\beta_{{\qo}'}=\bar\beta_{\qo}\setminus\beta_1$. Then, using the recursion properties of the DWPF we obtain
\begin{multline}\label{G2-pole}
\Bigl. \Lambda^{(l)}_m(\bar\alpha|\bar\beta)\Bigr|_{\alpha_1\to \beta_1}
={\sum}'  f(\bar\beta_{\qt},\bar\beta_{{\qo}'})f(\bar\beta_{\qt},\beta_{1})
 f(\bar\alpha_{{\qo}'},\bar\alpha_{\qt}) f(\alpha_{1},\bar\alpha_{\qt})\\
 \times
 g(\beta_1,\alpha_1)f(\alpha_{1},\bar\alpha_{{\qo}'})f(\bar\beta_{{\qo}'},\beta_{1})
 K_{k-1}(\bar\beta_{{\qo}'}|\bar\alpha_{{\qo}'})K_{m-k}(\bar\alpha_{\qt}+c|\bar\beta_{\qt}),
       \end{multline}
where ${\sum}'$ means that the sum is taken over partitions of the sets $\bar\alpha'$ and $\bar\beta'$. It is easy to see that
\be{comb}
\begin{array}{l}
f(\alpha_{1},\bar\alpha_{\qt})f(\alpha_{1},\bar\alpha_{{\qo}'})=f(\alpha_{1},\bar\alpha'),\\
f(\bar\beta_{\qt},\beta_{1})f(\bar\beta_{{\qo}'},\beta_{1})=f(\bar\beta',\beta_{1}),
\end{array}
\ee
and these factors can be moved out off the sum over partitions. We obtain
%
\begin{eqnarray}\label{G2-pole2}
\Bigl. \Lambda^{(l)}_m(\bar\alpha|\bar\beta)\Bigr|_{\alpha_1\to \beta_1}
&=&g(\beta_1,\alpha_1)f(\alpha_{1},\bar\alpha')f(\bar\beta',\beta_{1})
\nonumber \\
&& \times {\sum}'  f(\bar\beta_{\qt},\bar\beta_{{\qo}'})
 f(\bar\alpha_{{\qo}'},\bar\alpha_{\qt})
  K_{k-1}(\bar\beta_{{\qo}'}|\bar\alpha_{{\qo}'})K_{m-k}(\bar\alpha_{\qt}+c|\bar\beta_{\qt})
  \nonumber\\
 & =&-g(\alpha_1,\beta_1)f(\alpha_{1},\bar\alpha')f(\bar\beta',\beta_{1})
  \cdot \Lambda^{(l)}_{m-1}(\bar\alpha'|\bar\beta').
\end{eqnarray}
Similarly one can prove that
\be{sec-rec}
\Bigl. \Lambda^{(l)}_m(\bar\alpha|\bar\beta)\Bigr|_{\alpha_1+c \to \beta_1} =
 h^{-1}(\alpha_1, \beta_1)f(\beta_1,\bar\beta')f(\bar\alpha',\alpha_1)
 \cdot \Lambda^{(l)}_{m-1}(\bar\alpha'|\bar\beta').
\ee
Then the induction over $m$ ends the proof.\hfill $\square$

\end{document}